\documentclass[longauth]{aa}
\renewcommand{\eprint}[2][]{%
  \if!#1!#2\else#1\arxivprefixesep\ignorespaces#2\fi
}

\usepackage{graphicx}
\usepackage{natbib}
\usepackage{scalerel}
\usepackage{comment}
\usepackage{multirow}
\usepackage{ulem} 

\usepackage[table]{xcolor}

\usepackage{txfonts}
\usepackage[pdfencoding=auto,psdextra]{hyperref}
\hypersetup{
    colorlinks=true,
    linkcolor=blue,
    filecolor=magenta,      
    urlcolor=blue,
    citecolor=blue
}
\newcommand {\Sec}{Sect.\,}
\newcommand {\Fig}{Fig.\,}

\makeatletter
\renewcommand*\aa@pageof{, page \thepage{} of \pageref*{LastPage}}
\makeatother

\usepackage[utf8]{inputenc}

\usepackage{cuted}
\usepackage{capt-of}
\usepackage[switch, modulo]{lineno}
\let\linenumbers\relax

\usepackage{siunitx}
\usepackage{amsmath}
\usepackage{amssymb}
\usepackage{xspace}
\usepackage{euclid}
\usepackage{orcidlink}
\let\orcid\orcidlink

\begin{document}

\title{Euclid Data Release 1 (DR1)}
\subtitle{NISP spectroscopy of gravitational arcs and multiple images}
\author{Euclid Collaboration: P.~Bergamini\orcid{0000-0003-1383-9414}\thanks{\email{pietro.bergamini@inaf.it}}\inst{\ref{aff1},\ref{aff2}}
\and P.~Rosati\orcid{0000-0002-6813-0632}\inst{\ref{aff3},\ref{aff1}}
\and M.~Meneghetti\orcid{0000-0003-1225-7084}\inst{\ref{aff1},\ref{aff4}}
\and M.~Melloni\inst{\ref{aff3}}
\and B.~R.~Granett\orcid{0000-0003-2694-9284}\inst{\ref{aff5}}
\and L.~Tortorelli\orcid{0000-0002-8012-7495}\inst{\ref{aff6}}
\and A.~Acebron\orcid{0000-0003-3108-9039}\inst{\ref{aff7},\ref{aff8}}
\and M.~Annunziatella\orcid{0000-0002-8053-8040}\inst{\ref{aff8}}
\and G.~Angora\orcid{0000-0002-0316-6562}\inst{\ref{aff9},\ref{aff3}}
\and M.~Fogliardi\orcid{0009-0006-4964-5311}\inst{\ref{aff3},\ref{aff1}}
\and M.~Fumana\orcid{0000-0001-6787-5950}\inst{\ref{aff8}}
\and M.~Scodeggio\inst{\ref{aff8}}
\and M.~Talia\orcid{0000-0003-4352-2063}\inst{\ref{aff10},\ref{aff1}}
\and M.~Bolzonella\orcid{0000-0003-3278-4607}\inst{\ref{aff1}}
\and A.~Enia\orcid{0000-0002-0200-2857}\inst{\ref{aff1}}
\and E.~Franceschi\orcid{0000-0002-0585-6591}\inst{\ref{aff1}}
\and C.~Grillo\orcid{0000-0002-5926-7143}\inst{\ref{aff2},\ref{aff8}}
\and L.~Bazzanini\orcid{0000-0003-0727-0137}\inst{\ref{aff3},\ref{aff1}}
\and S.~M.~Callegaro\orcid{0009-0002-1312-0100}\inst{\ref{aff3}}
\and G.~Di~Rosa\orcid{0009-0001-9416-0923}\inst{\ref{aff3}}
\and M.~D'Addona\orcid{0000-0003-3445-0483}\inst{\ref{aff9},\ref{aff11}}
\and L.~R.~Ecker\orcid{0009-0005-3508-2469}\inst{\ref{aff6},\ref{aff12},\ref{aff13}}
\and R.~Gavazzi\orcid{0000-0002-5540-6935}\inst{\ref{aff14},\ref{aff15}}
\and G.~Granata\orcid{0000-0002-9512-3788}\inst{\ref{aff16}}
\and N.~E.~P.~Lines\orcid{0009-0004-7751-1914}\inst{\ref{aff16}}
\and M.~Lombardi\orcid{0000-0002-3336-4965}\inst{\ref{aff2},\ref{aff1}}
\and J.~W.~Nightingale\orcid{0000-0002-8987-7401}\inst{\ref{aff17}}
\and S.~Schuldt\orcid{0000-0003-2497-6334}\inst{\ref{aff18},\ref{aff19},\ref{aff8}}
\and S.~Serjeant\orcid{0000-0002-0517-7943}\inst{\ref{aff20}}
\and C.~J.~Storfer\orcid{0000-0002-0385-0014}\inst{\ref{aff21}}
\and S.~H.~Vincken\orcid{0009-0005-7305-2359}\inst{\ref{aff22}}
\and B.~Altieri\orcid{0000-0003-3936-0284}\inst{\ref{aff23}}
\and S.~Andreon\orcid{0000-0002-2041-8784}\inst{\ref{aff5}}
\and N.~Auricchio\orcid{0000-0003-4444-8651}\inst{\ref{aff1}}
\and C.~Baccigalupi\orcid{0000-0002-8211-1630}\inst{\ref{aff24},\ref{aff25},\ref{aff26},\ref{aff27}}
\and M.~Baldi\orcid{0000-0003-4145-1943}\inst{\ref{aff28},\ref{aff1},\ref{aff4}}
\and A.~Balestra\orcid{0000-0002-6967-261X}\inst{\ref{aff29}}
\and S.~Bardelli\orcid{0000-0002-8900-0298}\inst{\ref{aff1}}
\and P.~Battaglia\orcid{0000-0002-7337-5909}\inst{\ref{aff1}}
\and A.~Biviano\orcid{0000-0002-0857-0732}\inst{\ref{aff25},\ref{aff24}}
\and E.~Branchini\orcid{0000-0002-0808-6908}\inst{\ref{aff30},\ref{aff31},\ref{aff5}}
\and M.~Brescia\orcid{0000-0001-9506-5680}\inst{\ref{aff32},\ref{aff9}}
\and J.~Brinchmann\orcid{0000-0003-4359-8797}\inst{\ref{aff33},\ref{aff34},\ref{aff35}}
\and S.~Camera\orcid{0000-0003-3399-3574}\inst{\ref{aff36},\ref{aff37},\ref{aff38}}
\and V.~Capobianco\orcid{0000-0002-3309-7692}\inst{\ref{aff38}}
\and C.~Carbone\orcid{0000-0003-0125-3563}\inst{\ref{aff8}}
\and J.~Carretero\orcid{0000-0002-3130-0204}\inst{\ref{aff39},\ref{aff40}}
\and M.~Castellano\orcid{0000-0001-9875-8263}\inst{\ref{aff41}}
\and G.~Castignani\orcid{0000-0001-6831-0687}\inst{\ref{aff1}}
\and S.~Cavuoti\orcid{0000-0002-3787-4196}\inst{\ref{aff9},\ref{aff42}}
\and K.~C.~Chambers\orcid{0000-0001-6965-7789}\inst{\ref{aff21}}
\and A.~Cimatti\inst{\ref{aff43}}
\and C.~Colodro-Conde\inst{\ref{aff44}}
\and G.~Congedo\orcid{0000-0003-2508-0046}\inst{\ref{aff45}}
\and C.~J.~Conselice\orcid{0000-0003-1949-7638}\inst{\ref{aff46}}
\and L.~Conversi\orcid{0000-0002-6710-8476}\inst{\ref{aff47},\ref{aff23}}
\and Y.~Copin\orcid{0000-0002-5317-7518}\inst{\ref{aff48}}
\and A.~Costille\inst{\ref{aff14}}
\and F.~Courbin\orcid{0000-0003-0758-6510}\inst{\ref{aff49},\ref{aff50},\ref{aff51}}
\and H.~M.~Courtois\orcid{0000-0003-0509-1776}\inst{\ref{aff52}}
\and J.-G.~Cuby\orcid{0000-0002-8767-1442}\inst{\ref{aff53},\ref{aff14}}
\and H.~Degaudenzi\orcid{0000-0002-5887-6799}\inst{\ref{aff54}}
\and S.~de~la~Torre\inst{\ref{aff14}}
\and G.~De~Lucia\orcid{0000-0002-6220-9104}\inst{\ref{aff25}}
\and H.~Dole\orcid{0000-0002-9767-3839}\inst{\ref{aff55}}
\and F.~Dubath\orcid{0000-0002-6533-2810}\inst{\ref{aff54}}
\and X.~Dupac\inst{\ref{aff23}}
\and S.~Dusini\orcid{0000-0002-1128-0664}\inst{\ref{aff56}}
\and S.~Escoffier\orcid{0000-0002-2847-7498}\inst{\ref{aff57}}
\and M.~Farina\orcid{0000-0002-3089-7846}\inst{\ref{aff58}}
\and R.~Farinelli\inst{\ref{aff1}}
\and S.~Ferriol\inst{\ref{aff48}}
\and F.~Finelli\orcid{0000-0002-6694-3269}\inst{\ref{aff1},\ref{aff59}}
\and N.~Fourmanoit\orcid{0009-0005-6816-6925}\inst{\ref{aff57}}
\and M.~Frailis\orcid{0000-0002-7400-2135}\inst{\ref{aff25}}
\and L.~Gabarra\orcid{0000-0002-8486-8856}\inst{\ref{aff60}}
\and S.~Galeotta\orcid{0000-0002-3748-5115}\inst{\ref{aff25}}
\and K.~George\orcid{0000-0002-1734-8455}\inst{\ref{aff61}}
\and B.~Gillis\orcid{0000-0002-4478-1270}\inst{\ref{aff45}}
\and C.~Giocoli\orcid{0000-0002-9590-7961}\inst{\ref{aff1},\ref{aff4}}
\and J.~Gracia-Carpio\orcid{0000-0003-4689-3134}\inst{\ref{aff12}}
\and A.~Grazian\orcid{0000-0002-5688-0663}\inst{\ref{aff29}}
\and F.~Grupp\inst{\ref{aff12},\ref{aff6}}
\and L.~Guzzo\orcid{0000-0001-8264-5192}\inst{\ref{aff2},\ref{aff5},\ref{aff62}}
\and S.~Gwyn\orcid{0000-0001-8221-8406}\inst{\ref{aff63}}
\and W.~G.~Hartley\orcid{0000-0002-4563-5101}\inst{\ref{aff54}}
\and S.~V.~H.~Haugan\orcid{0000-0001-9648-7260}\inst{\ref{aff64}}
\and G.~Helou\orcid{0000-0003-3367-3415}\inst{\ref{aff65},\ref{aff66}}
\and S.~Hemmati\orcid{0000-0003-2226-5395}\inst{\ref{aff66}}
\and J.~Hoar\inst{\ref{aff23}}
\and H.~Hoekstra\orcid{0000-0002-0641-3231}\inst{\ref{aff67}}
\and W.~Holmes\orcid{0009-0007-8554-4646}\inst{\ref{aff68}}
\and F.~Hormuth\inst{\ref{aff69}}
\and A.~Hornstrup\orcid{0000-0002-3363-0936}\inst{\ref{aff70},\ref{aff71}}
\and K.~Jahnke\orcid{0000-0003-3804-2137}\inst{\ref{aff72}}
\and M.~Jhabvala\inst{\ref{aff73}}
\and S.~Kermiche\orcid{0000-0002-0302-5735}\inst{\ref{aff57}}
\and A.~Kiessling\orcid{0000-0002-2590-1273}\inst{\ref{aff68}}
\and B.~Kubik\orcid{0009-0006-5823-4880}\inst{\ref{aff48}}
\and M.~K\"ummel\orcid{0000-0003-2791-2117}\inst{\ref{aff6}}
\and M.~Kunz\orcid{0000-0002-3052-7394}\inst{\ref{aff74}}
\and H.~Kurki-Suonio\orcid{0000-0002-4618-3063}\inst{\ref{aff19},\ref{aff75}}
\and A.~M.~C.~Le~Brun\orcid{0000-0002-0936-4594}\inst{\ref{aff76}}
\and V.~Le~Brun\orcid{0000-0002-5027-1939}\inst{\ref{aff14}}
\and D.~Le~Mignant\orcid{0000-0002-5339-5515}\inst{\ref{aff14}}
\and S.~Ligori\orcid{0000-0003-4172-4606}\inst{\ref{aff38}}
\and P.~B.~Lilje\orcid{0000-0003-4324-7794}\inst{\ref{aff64}}
\and V.~Lindholm\orcid{0000-0003-2317-5471}\inst{\ref{aff19},\ref{aff75}}
\and I.~Lloro\orcid{0000-0001-5966-1434}\inst{\ref{aff77}}
\and M.~Magliocchetti\orcid{0000-0001-9158-4838}\inst{\ref{aff58}}
\and G.~Mainetti\orcid{0000-0003-2384-2377}\inst{\ref{aff78}}
\and O.~Mansutti\orcid{0000-0001-5758-4658}\inst{\ref{aff25}}
\and S.~Marcin\inst{\ref{aff79}}
\and O.~Marggraf\orcid{0000-0001-7242-3852}\inst{\ref{aff80}}
\and M.~Martinelli\orcid{0000-0002-6943-7732}\inst{\ref{aff41},\ref{aff81}}
\and N.~Martinet\orcid{0000-0003-2786-7790}\inst{\ref{aff14}}
\and F.~Marulli\orcid{0000-0002-8850-0303}\inst{\ref{aff10},\ref{aff1},\ref{aff4}}
\and R.~J.~Massey\orcid{0000-0002-6085-3780}\inst{\ref{aff82}}
\and N.~Mauri\orcid{0000-0001-8196-1548}\inst{\ref{aff43},\ref{aff4}}
\and E.~Medinaceli\orcid{0000-0002-4040-7783}\inst{\ref{aff1}}
\and S.~Mei\orcid{0000-0002-2849-559X}\inst{\ref{aff83},\ref{aff84}}
\and E.~Merlin\orcid{0000-0001-6870-8900}\inst{\ref{aff29}}
\and G.~Meylan\orcid{0000-0001-6503-0209}\inst{\ref{aff85}}
\and J.~J.~Mohr\orcid{0000-0002-6875-2087}\inst{\ref{aff61}}
\and P.~Monaco\orcid{0000-0003-2083-7564}\inst{\ref{aff86},\ref{aff25},\ref{aff26},\ref{aff24}}
\and A.~Mora\orcid{0000-0002-1922-8529}\inst{\ref{aff87}}
\and M.~Moresco\orcid{0000-0002-7616-7136}\inst{\ref{aff10},\ref{aff1}}
\and C.~Moretti\orcid{0000-0003-3314-8936}\inst{\ref{aff25},\ref{aff24},\ref{aff26}}
\and L.~Moscardini\orcid{0000-0002-3473-6716}\inst{\ref{aff10},\ref{aff1},\ref{aff4}}
\and E.~Munari\orcid{0000-0002-1751-5946}\inst{\ref{aff25},\ref{aff24}}
\and C.~Neissner\orcid{0000-0001-8524-4968}\inst{\ref{aff88},\ref{aff40}}
\and R.~C.~Nichol\orcid{0000-0003-0939-6518}\inst{\ref{aff89}}
\and S.-M.~Niemi\orcid{0009-0005-0247-0086}\inst{\ref{aff90}}
\and C.~Padilla\orcid{0000-0001-7951-0166}\inst{\ref{aff88}}
\and S.~Paltani\orcid{0000-0002-8108-9179}\inst{\ref{aff54}}
\and F.~Pasian\orcid{0000-0002-4869-3227}\inst{\ref{aff25}}
\and W.~J.~Percival\orcid{0000-0002-0644-5727}\inst{\ref{aff91},\ref{aff92},\ref{aff93}}
\and V.~Pettorino\orcid{0000-0002-4203-9320}\inst{\ref{aff90}}
\and A.~Pezzotta\orcid{0000-0003-0726-2268}\inst{\ref{aff5}}
\and S.~Pires\orcid{0000-0002-0249-2104}\inst{\ref{aff94}}
\and G.~Polenta\orcid{0000-0003-4067-9196}\inst{\ref{aff95}}
\and M.~Poncet\inst{\ref{aff96}}
\and L.~A.~Popa\inst{\ref{aff97}}
\and L.~Pozzetti\orcid{0000-0001-7085-0412}\inst{\ref{aff1}}
\and F.~Raison\orcid{0000-0002-7819-6918}\inst{\ref{aff12}}
\and R.~Rebolo\orcid{0000-0003-3767-7085}\inst{\ref{aff44},\ref{aff98},\ref{aff99}}
\and A.~Renzi\orcid{0000-0001-9856-1970}\inst{\ref{aff100},\ref{aff56},\ref{aff1}}
\and J.~Rhodes\orcid{0000-0002-4485-8549}\inst{\ref{aff68}}
\and G.~Riccio\inst{\ref{aff9}}
\and I.~Risso\orcid{0000-0003-2525-7761}\inst{\ref{aff30},\ref{aff31},\ref{aff5}}
\and F.~Rizzo\orcid{0000-0002-9407-585X}\inst{\ref{aff25}}
\and E.~Romelli\orcid{0000-0003-3069-9222}\inst{\ref{aff25}}
\and M.~Roncarelli\orcid{0000-0001-9587-7822}\inst{\ref{aff1}}
\and B.~Rusholme\orcid{0000-0001-7648-4142}\inst{\ref{aff66}}
\and R.~Saglia\orcid{0000-0003-0378-7032}\inst{\ref{aff6},\ref{aff12}}
\and Z.~Sakr\orcid{0000-0002-4823-3757}\inst{\ref{aff101},\ref{aff102},\ref{aff103}}
\and D.~Sapone\orcid{0000-0001-7089-4503}\inst{\ref{aff104}}
\and B.~Sartoris\orcid{0000-0003-1337-5269}\inst{\ref{aff12},\ref{aff25}}
\and P.~Schneider\orcid{0000-0001-8561-2679}\inst{\ref{aff80}}
\and T.~Schrabback\orcid{0000-0002-6987-7834}\inst{\ref{aff105}}
\and A.~Secroun\orcid{0000-0003-0505-3710}\inst{\ref{aff57}}
\and E.~Sihvola\orcid{0000-0003-1804-7715}\inst{\ref{aff106}}
\and P.~Simon\inst{\ref{aff80}}
\and C.~Sirignano\orcid{0000-0002-0995-7146}\inst{\ref{aff100},\ref{aff56}}
\and G.~Sirri\orcid{0000-0003-2626-2853}\inst{\ref{aff4}}
\and L.~Stanco\orcid{0000-0002-9706-5104}\inst{\ref{aff56}}
\and C.~Surace\orcid{0000-0003-2592-0113}\inst{\ref{aff14}}
\and P.~Tallada-Cresp\'{i}\orcid{0000-0002-1336-8328}\inst{\ref{aff39},\ref{aff40}}
\and A.~N.~Taylor\inst{\ref{aff45}}
\and I.~Tereno\orcid{0000-0002-4537-6218}\inst{\ref{aff107},\ref{aff108}}
\and N.~Tessore\orcid{0000-0002-9696-7931}\inst{\ref{aff109}}
\and S.~Toft\orcid{0000-0003-3631-7176}\inst{\ref{aff110},\ref{aff111}}
\and R.~Toledo-Moreo\orcid{0000-0002-2997-4859}\inst{\ref{aff112},\ref{aff113}}
\and F.~Torradeflot\orcid{0000-0003-1160-1517}\inst{\ref{aff40},\ref{aff39}}
\and A.~Tsyganov\inst{\ref{aff114}}
\and I.~Tutusaus\orcid{0000-0002-3199-0399}\inst{\ref{aff115},\ref{aff116},\ref{aff102}}
\and J.~Valiviita\orcid{0000-0001-6225-3693}\inst{\ref{aff19},\ref{aff75}}
\and T.~Vassallo\orcid{0000-0001-6512-6358}\inst{\ref{aff25},\ref{aff61}}
\and A.~Veropalumbo\orcid{0000-0003-2387-1194}\inst{\ref{aff5},\ref{aff31},\ref{aff30}}
\and Y.~Wang\orcid{0000-0002-4749-2984}\inst{\ref{aff66}}
\and J.~Weller\orcid{0000-0002-8282-2010}\inst{\ref{aff6},\ref{aff12}}
\and A.~Zacchei\orcid{0000-0003-0396-1192}\inst{\ref{aff25},\ref{aff24}}
\and G.~Zamorani\orcid{0000-0002-2318-301X}\inst{\ref{aff1}}
\and E.~Zucca\orcid{0000-0002-5845-8132}\inst{\ref{aff1}}
\and M.~Ballardini\orcid{0000-0003-4481-3559}\inst{\ref{aff3},\ref{aff117},\ref{aff1}}
\and S.~Borgani\orcid{0000-0001-6151-6439}\inst{\ref{aff86},\ref{aff24},\ref{aff25},\ref{aff26},\ref{aff118}}
\and E.~Bozzo\orcid{0000-0002-8201-1525}\inst{\ref{aff54}}
\and C.~Burigana\orcid{0000-0002-3005-5796}\inst{\ref{aff119},\ref{aff59}}
\and R.~Cabanac\orcid{0000-0001-6679-2600}\inst{\ref{aff102}}
\and A.~Cappi\inst{\ref{aff120},\ref{aff1}}
\and F.~Caro\orcid{0009-0003-1053-0507}\inst{\ref{aff41}}
\and T.~Castro\orcid{0000-0002-6292-3228}\inst{\ref{aff121},\ref{aff25},\ref{aff24}}
\and D.~Di~Ferdinando\inst{\ref{aff4}}
\and J.~A.~Escartin~Vigo\inst{\ref{aff12}}
\and J.~Garc\'ia-Bellido\orcid{0000-0002-9370-8360}\inst{\ref{aff101}}
\and T.~Gasparetto\orcid{0000-0002-7913-4866}\inst{\ref{aff41}}
\and C.~M.~Gutierrez\orcid{0000-0001-7854-783X}\inst{\ref{aff44},\ref{aff99}}
\and A.~Loureiro\orcid{0000-0002-4371-0876}\inst{\ref{aff122},\ref{aff123}}
\and J.~Macias-Perez\orcid{0000-0002-5385-2763}\inst{\ref{aff124}}
\and R.~Maoli\orcid{0000-0002-6065-3025}\inst{\ref{aff125},\ref{aff41}}
\and R.~B.~Metcalf\orcid{0000-0003-3167-2574}\inst{\ref{aff10},\ref{aff1}}
\and A.~A.~Nucita\inst{\ref{aff126},\ref{aff127},\ref{aff128}}
\and M.~P\"ontinen\orcid{0000-0001-5442-2530}\inst{\ref{aff19}}
\and E.~Sarpa\orcid{0000-0002-1256-655X}\inst{\ref{aff25}}
\and V.~Scottez\orcid{0009-0008-3864-940X}\inst{\ref{aff129},\ref{aff130}}
\and M.~Sereno\orcid{0000-0003-0302-0325}\inst{\ref{aff1},\ref{aff4}}
\and M.~Tenti\orcid{0000-0002-4254-5901}\inst{\ref{aff4}}
\and M.~Tucci\inst{\ref{aff54}}
\and M.~Viel\orcid{0000-0002-2642-5707}\inst{\ref{aff24},\ref{aff25},\ref{aff27},\ref{aff26},\ref{aff118}}
\and M.~Wiesmann\orcid{0009-0000-8199-5860}\inst{\ref{aff64}}
\and J.~A.~Acevedo~Barroso\orcid{0000-0002-9654-1711}\inst{\ref{aff68}}
\and Y.~Akrami\orcid{0000-0002-2407-7956}\inst{\ref{aff101},\ref{aff131}}
\and I.~T.~Andika\orcid{0000-0001-6102-9526}\inst{\ref{aff6}}
\and S.~Anselmi\orcid{0000-0002-3579-9583}\inst{\ref{aff56},\ref{aff100},\ref{aff132}}
\and M.~Archidiacono\orcid{0000-0003-4952-9012}\inst{\ref{aff2},\ref{aff62}}
\and G.~Aric\`o\orcid{0000-0002-2802-2928}\inst{\ref{aff4}}
\and F.~Atrio-Barandela\orcid{0000-0002-2130-2513}\inst{\ref{aff133}}
\and M.~Baes\orcid{0000-0002-3930-2757}\inst{\ref{aff134}}
\and D.~Bertacca\orcid{0000-0002-2490-7139}\inst{\ref{aff100},\ref{aff29},\ref{aff56}}
\and M.~Bethermin\orcid{0000-0002-3915-2015}\inst{\ref{aff135}}
\and F.~Beutler\orcid{0000-0003-0467-5438}\inst{\ref{aff45}}
\and L.~Blot\orcid{0000-0002-9622-7167}\inst{\ref{aff136},\ref{aff76}}
\and H.~B\"ohringer\orcid{0000-0001-8241-4204}\inst{\ref{aff12},\ref{aff61},\ref{aff137}}
\and M.~L.~Brown\orcid{0000-0002-0370-8077}\inst{\ref{aff46}}
\and S.~Bruton\orcid{0000-0002-6503-5218}\inst{\ref{aff65}}
\and A.~Calabro\orcid{0000-0003-2536-1614}\inst{\ref{aff41}}
\and B.~Camacho~Quevedo\orcid{0000-0002-8789-4232}\inst{\ref{aff24},\ref{aff27},\ref{aff25}}
\and C.~S.~Carvalho\inst{\ref{aff108}}
\and Y.~Charles\orcid{0009-0000-3636-834X}\inst{\ref{aff14}}
\and F.~Cogato\orcid{0000-0003-4632-6113}\inst{\ref{aff10},\ref{aff1}}
\and T.~E.~Collett\orcid{0000-0001-5564-3140}\inst{\ref{aff16}}
\and S.~Conseil\orcid{0000-0002-3657-4191}\inst{\ref{aff138},\ref{aff48}}
\and A.~R.~Cooray\orcid{0000-0002-3892-0190}\inst{\ref{aff139}}
\and P.~Corcho-Caballero\orcid{0000-0001-6327-7080}\inst{\ref{aff140}}
\and B.~Csizi\orcid{0000-0003-3227-6581}\inst{\ref{aff105}}
\and O.~Cucciati\orcid{0000-0002-9336-7551}\inst{\ref{aff1}}
\and H.~Dannerbauer\orcid{0000-0001-7147-3575}\inst{\ref{aff44},\ref{aff99}}
\and T.~de~Boer\orcid{0000-0001-5486-2747}\inst{\ref{aff21}}
\and G.~Desprez\orcid{0000-0001-8325-1742}\inst{\ref{aff140}}
\and A.~D\'iaz-S\'anchez\orcid{0000-0003-0748-4768}\inst{\ref{aff141}}
\and S.~Di~Domizio\orcid{0000-0003-2863-5895}\inst{\ref{aff30},\ref{aff31}}
\and J.~M.~Diego\orcid{0000-0001-9065-3926}\inst{\ref{aff7}}
\and P.-A.~Duc\orcid{0000-0003-3343-6284}\inst{\ref{aff135}}
\and V.~Duret\orcid{0009-0009-0383-4960}\inst{\ref{aff57}}
\and M.~Y.~Elkhashab\orcid{0000-0001-9306-2603}\inst{\ref{aff25},\ref{aff26},\ref{aff86},\ref{aff24}}
\and Y.~Fang\orcid{0000-0002-0334-6950}\inst{\ref{aff6}}
\and A.~Farina\orcid{0009-0000-3420-929X}\inst{\ref{aff5},\ref{aff31}}
\and A.~Finoguenov\orcid{0000-0002-4606-5403}\inst{\ref{aff19}}
\and A.~Franco\orcid{0000-0002-4761-366X}\inst{\ref{aff126},\ref{aff127},\ref{aff128}}
\and Y.~Fu\orcid{0000-0002-0759-0504}\inst{\ref{aff67},\ref{aff140}}
\and K.~Ganga\orcid{0000-0001-8159-8208}\inst{\ref{aff83}}
\and E.~Gaztanaga\orcid{0000-0001-9632-0815}\inst{\ref{aff115},\ref{aff116},\ref{aff16}}
\and Z.~Ghaffari\orcid{0000-0002-6467-8078}\inst{\ref{aff25},\ref{aff24}}
\and F.~Giacomini\orcid{0000-0002-3129-2814}\inst{\ref{aff4}}
\and F.~Gianotti\orcid{0000-0003-4666-119X}\inst{\ref{aff1}}
\and G.~Gozaliasl\orcid{0000-0002-0236-919X}\inst{\ref{aff142},\ref{aff19}}
\and A.~Gruppuso\orcid{0000-0001-9272-5292}\inst{\ref{aff1},\ref{aff4}}
\and M.~Guidi\orcid{0000-0001-9408-1101}\inst{\ref{aff28},\ref{aff1}}
\and A.~Hall\orcid{0000-0002-3139-8651}\inst{\ref{aff45}}
\and N.~A.~Hatch\orcid{0000-0001-5600-0534}\inst{\ref{aff143}}
\and T.~Heuberger\orcid{0009-0009-3474-8467}\inst{\ref{aff79}}
\and H.~Hildebrandt\orcid{0000-0002-9814-3338}\inst{\ref{aff144}}
\and J.~Hjorth\orcid{0000-0002-4571-2306}\inst{\ref{aff145}}
\and M.~Jauzac\orcid{0000-0003-1974-8732}\inst{\ref{aff146},\ref{aff82},\ref{aff147},\ref{aff148},\ref{aff102}}
\and J.~J.~E.~Kajava\orcid{0000-0002-3010-8333}\inst{\ref{aff149},\ref{aff18},\ref{aff150}}
\and Y.~Kang\orcid{0009-0000-8588-7250}\inst{\ref{aff105}}
\and V.~Kansal\orcid{0000-0002-4008-6078}\inst{\ref{aff151},\ref{aff152}}
\and D.~Karagiannis\orcid{0000-0002-4927-0816}\inst{\ref{aff3},\ref{aff153}}
\and J.~Kim\orcid{0000-0003-2776-2761}\inst{\ref{aff60}}
\and C.~C.~Kirkpatrick\inst{\ref{aff106}}
\and A.~Kov\'acs\orcid{0000-0002-5825-579X}\inst{\ref{aff154},\ref{aff155}}
\and I.~Kova{\v{c}}i{\'{c}}\orcid{0000-0001-6751-3263}\inst{\ref{aff134}}
\and K.~Koyama\orcid{0000-0001-6727-6915}\inst{\ref{aff16}}
\and S.~Kruk\orcid{0000-0001-8010-8879}\inst{\ref{aff23}}
\and M.~C.~Lam\orcid{0000-0002-9347-2298}\inst{\ref{aff45}}
\and M.~Lattanzi\orcid{0000-0003-1059-2532}\inst{\ref{aff117}}
\and G.~Leroy\orcid{0009-0004-2523-4425}\inst{\ref{aff146},\ref{aff82}}
\and G.~F.~Lesci\orcid{0000-0002-4607-2830}\inst{\ref{aff10},\ref{aff1}}
\and J.~Lesgourgues\orcid{0000-0001-7627-353X}\inst{\ref{aff156}}
\and L.~Linke\orcid{0000-0002-2622-8113}\inst{\ref{aff105}}
\and G.~Maggio\orcid{0000-0003-4020-4836}\inst{\ref{aff25}}
\and E.~A.~Magnier\orcid{0000-0002-7965-2815}\inst{\ref{aff21}}
\and F.~Mannucci\orcid{0000-0002-4803-2381}\inst{\ref{aff157}}
\and F.~R.~Marleau\orcid{0000-0002-1442-2947}\inst{\ref{aff105}}
\and C.~J.~A.~P.~Martins\orcid{0000-0002-4886-9261}\inst{\ref{aff158},\ref{aff33}}
\and M.~Miluzio\inst{\ref{aff23},\ref{aff159}}
\and G.~Morgante\inst{\ref{aff1}}
\and K.~Naidoo\orcid{0000-0002-9182-1802}\inst{\ref{aff16},\ref{aff72}}
\and P.~Natoli\orcid{0000-0003-0126-9100}\inst{\ref{aff3},\ref{aff117}}
\and A.~Navarro-Alsina\orcid{0000-0002-3173-2592}\inst{\ref{aff80}}
\and A.~Nersesian\orcid{0000-0001-6843-409X}\inst{\ref{aff160},\ref{aff134}}
\and S.~Nesseris\orcid{0000-0002-0567-0324}\inst{\ref{aff101}}
\and L.~Nicastro\orcid{0000-0001-8534-6788}\inst{\ref{aff1}}
\and F.~Oppizzi\orcid{0000-0003-3904-8370}\inst{\ref{aff31},\ref{aff56},\ref{aff100}}
\and F.~Pace\orcid{0000-0001-8039-0480}\inst{\ref{aff36},\ref{aff37},\ref{aff38},\ref{aff161}}
\and E.~Palazzi\orcid{0000-0002-8691-7666}\inst{\ref{aff1}}
\and D.~Paoletti\orcid{0000-0003-4761-6147}\inst{\ref{aff1},\ref{aff59}}
\and F.~Passalacqua\orcid{0000-0002-8606-4093}\inst{\ref{aff56}}
\and K.~Paterson\orcid{0000-0001-8340-3486}\inst{\ref{aff72}}
\and C.~Pattison\orcid{0000-0003-3272-2617}\inst{\ref{aff16}}
\and A.~Pisani\orcid{0000-0002-6146-4437}\inst{\ref{aff57}}
\and D.~Potter\orcid{0000-0002-0757-5195}\inst{\ref{aff162}}
\and G.~W.~Pratt\inst{\ref{aff94}}
\and S.~Quai\orcid{0000-0002-0449-8163}\inst{\ref{aff10},\ref{aff1}}
\and M.~Radovich\orcid{0000-0002-3585-866X}\inst{\ref{aff29}}
\and F.~Ricci\orcid{0000-0001-5742-5980}\inst{\ref{aff163},\ref{aff41}}
\and G.~Rodighiero\orcid{0000-0002-9415-2296}\inst{\ref{aff100},\ref{aff29}}
\and W.~Roster\orcid{0000-0002-9149-6528}\inst{\ref{aff12}}
\and M.~Sahl\'en\orcid{0000-0003-0973-4804}\inst{\ref{aff164}}
\and D.~B.~Sanders\orcid{0000-0002-1233-9998}\inst{\ref{aff21}}
\and A.~Schneider\orcid{0000-0001-7055-8104}\inst{\ref{aff162}}
\and D.~Sciotti\orcid{0009-0008-4519-2620}\inst{\ref{aff41},\ref{aff81}}
\and E.~Sellentin\orcid{0009-0002-2655-3458}\inst{\ref{aff165},\ref{aff67}}
\and F.~Shankar\orcid{0000-0001-8973-5051}\inst{\ref{aff166}}
\and L.~C.~Smith\orcid{0000-0002-3259-2771}\inst{\ref{aff167}}
\and J.~G.~Sorce\orcid{0000-0002-2307-2432}\inst{\ref{aff168},\ref{aff55}}
\and S.~A.~Stanford\orcid{0000-0003-0122-0841}\inst{\ref{aff169}}
\and I.~Szapudi\orcid{0000-0003-2274-0301}\inst{\ref{aff21}}
\and K.~Tanidis\orcid{0000-0001-9843-5130}\inst{\ref{aff170}}
\and C.~Tao\orcid{0000-0001-7961-8177}\inst{\ref{aff57}}
\and G.~Testera\orcid{0000-0003-2970-766X}\inst{\ref{aff31}}
\and R.~Teyssier\orcid{0000-0001-7689-0933}\inst{\ref{aff171}}
\and A.~Troja\orcid{0000-0003-0239-4595}\inst{\ref{aff25}}
\and C.~Uhlemann\orcid{0000-0001-7831-1579}\inst{\ref{aff172},\ref{aff17}}
\and C.~Valieri\inst{\ref{aff4}}
\and A.~Venhola\orcid{0000-0001-6071-4564}\inst{\ref{aff173}}
\and D.~Vergani\orcid{0000-0003-0898-2216}\inst{\ref{aff1}}
\and G.~Vernardos\orcid{0000-0001-8554-7248}\inst{\ref{aff174},\ref{aff175}}
\and P.~Vielzeuf\orcid{0000-0003-2035-9339}\inst{\ref{aff57}}
\and E.~Vilenius\orcid{0000-0002-6184-7681}\inst{\ref{aff109}}
\and S.~Vinciguerra\orcid{0009-0005-4018-3184}\inst{\ref{aff14}}
\and M.~von~Wietersheim-Kramsta\orcid{0000-0003-4986-5091}\inst{\ref{aff82},\ref{aff146}}
\and N.~A.~Walton\orcid{0000-0003-3983-8778}\inst{\ref{aff167}}
\and L.~Wang\orcid{0000-0002-6736-9158}\inst{\ref{aff176},\ref{aff140}}
\and A.~H.~Wright\orcid{0000-0001-7363-7932}\inst{\ref{aff144}}
\and H.~W.~Yeung\orcid{0000-0002-4993-9014}\inst{\ref{aff45}}
\and D.~Scott\orcid{0000-0002-6878-9840}\inst{\ref{aff177}}}
										   
\institute{INAF-Osservatorio di Astrofisica e Scienza dello Spazio di Bologna, Via Piero Gobetti 93/3, 40129 Bologna, Italy\label{aff1}
\and
Dipartimento di Fisica "Aldo Pontremoli", Universit\`a degli Studi di Milano, Via Celoria 16, 20133 Milano, Italy\label{aff2}
\and
Dipartimento di Fisica e Scienze della Terra, Universit\`a degli Studi di Ferrara, Via Giuseppe Saragat 1, 44122 Ferrara, Italy\label{aff3}
\and
INFN-Sezione di Bologna, Viale Berti Pichat 6/2, 40127 Bologna, Italy\label{aff4}
\and
INAF-Osservatorio Astronomico di Brera, Via Brera 28, 20122 Milano, Italy\label{aff5}
\and
Universit\"ats-Sternwarte M\"unchen, Fakult\"at f\"ur Physik, Ludwig-Maximilians-Universit\"at M\"unchen, Scheinerstr.~1, 81679 M\"unchen, Germany\label{aff6}
\and
Instituto de F\'isica de Cantabria, Edificio Juan Jord\'a, Avenida de los Castros, 39005 Santander, Spain\label{aff7}
\and
INAF-IASF Milano, Via Alfonso Corti 12, 20133 Milano, Italy\label{aff8}
\and
INAF-Osservatorio Astronomico di Capodimonte, Via Moiariello 16, 80131 Napoli, Italy\label{aff9}
\and
Dipartimento di Fisica e Astronomia "Augusto Righi" - Alma Mater Studiorum Universit\`a di Bologna, via Piero Gobetti 93/2, 40129 Bologna, Italy\label{aff10}
\and
Universita di Salerno, Dipartimento di Fisica "E.R. Caianiello", Via Giovanni Paolo II 132, I-84084 Fisciano (SA), Italy\label{aff11}
\and
Max Planck Institute for Extraterrestrial Physics, Giessenbachstr. 1, 85748 Garching, Germany\label{aff12}
\and
Max-Planck-Institut f\"ur Astrophysik, Karl-Schwarzschild-Str.~1, 85748 Garching, Germany\label{aff13}
\and
Aix-Marseille Universit\'e, CNRS, CNES, LAM, Marseille, France\label{aff14}
\and
Institut d'Astrophysique de Paris, UMR 7095, CNRS, and Sorbonne Universit\'e, 98 bis boulevard Arago, 75014 Paris, France\label{aff15}
\and
Institute of Cosmology and Gravitation, University of Portsmouth, Portsmouth PO1 3FX, UK\label{aff16}
\and
School of Mathematics, Statistics and Physics, Newcastle University, Herschel Building, Newcastle-upon-Tyne, NE1 7RU, UK\label{aff17}
\and
Finnish Centre for Astronomy with ESO (FINCA), Quantum, Vesilinnantie 5, University of Turku, 20014 Turku, Finland\label{aff18}
\and
Department of Physics, P.O. Box 64, University of Helsinki, 00014 Helsinki, Finland\label{aff19}
\and
School of Physical Sciences, The Open University, Milton Keynes, MK7 6AA, UK\label{aff20}
\and
Institute for Astronomy, University of Hawaii, 2680 Woodlawn Drive, Honolulu, HI 96822, USA\label{aff21}
\and
University of Applied Sciences and Arts of Northwestern Switzerland, School of Engineering, 5210 Windisch, Switzerland\label{aff22}
\and
ESAC/ESA, Camino Bajo del Castillo, s/n., Urb. Villafranca del Castillo, 28692 Villanueva de la Ca\~nada, Madrid, Spain\label{aff23}
\and
IFPU, Institute for Fundamental Physics of the Universe, via Beirut 2, 34151 Trieste, Italy\label{aff24}
\and
INAF-Osservatorio Astronomico di Trieste, Via G. B. Tiepolo 11, 34143 Trieste, Italy\label{aff25}
\and
INFN, Sezione di Trieste, Via Valerio 2, 34127 Trieste TS, Italy\label{aff26}
\and
SISSA, International School for Advanced Studies, Via Bonomea 265, 34136 Trieste TS, Italy\label{aff27}
\and
Dipartimento di Fisica e Astronomia, Universit\`a di Bologna, Via Gobetti 93/2, 40129 Bologna, Italy\label{aff28}
\and
INAF-Osservatorio Astronomico di Padova, Via dell'Osservatorio 5, 35122 Padova, Italy\label{aff29}
\and
Dipartimento di Fisica, Universit\`a di Genova, Via Dodecaneso 33, 16146, Genova, Italy\label{aff30}
\and
INFN-Sezione di Genova, Via Dodecaneso 33, 16146, Genova, Italy\label{aff31}
\and
Department of Physics "E. Pancini", University Federico II, Via Cinthia 6, 80126, Napoli, Italy\label{aff32}
\and
Instituto de Astrof\'isica e Ci\^encias do Espa\c{c}o, Universidade do Porto, CAUP, Rua das Estrelas, PT4150-762 Porto, Portugal\label{aff33}
\and
Faculdade de Ci\^encias da Universidade do Porto, Rua do Campo de Alegre, 4150-007 Porto, Portugal\label{aff34}
\and
European Southern Observatory, Karl-Schwarzschild-Str.~2, 85748 Garching, Germany\label{aff35}
\and
Dipartimento di Fisica, Universit\`a degli Studi di Torino, Via P. Giuria 1, 10125 Torino, Italy\label{aff36}
\and
INFN-Sezione di Torino, Via P. Giuria 1, 10125 Torino, Italy\label{aff37}
\and
INAF-Osservatorio Astrofisico di Torino, Via Osservatorio 20, 10025 Pino Torinese (TO), Italy\label{aff38}
\and
Centro de Investigaciones Energ\'eticas, Medioambientales y Tecnol\'ogicas (CIEMAT), Avenida Complutense 40, 28040 Madrid, Spain\label{aff39}
\and
Port d'Informaci\'{o} Cient\'{i}fica, Campus UAB, C. Albareda s/n, 08193 Bellaterra (Barcelona), Spain\label{aff40}
\and
INAF-Osservatorio Astronomico di Roma, Via Frascati 33, 00078 Monteporzio Catone, Italy\label{aff41}
\and
INFN -- Sezione di Napoli, Via Cinthia 6, 80126, Napoli, Italy\label{aff42}
\and
Dipartimento di Fisica e Astronomia "Augusto Righi" - Alma Mater Studiorum Universit\`a di Bologna, Viale Berti Pichat 6/2, 40127 Bologna, Italy\label{aff43}
\and
Instituto de Astrof\'{\i}sica de Canarias, E-38205 La Laguna, Tenerife, Spain\label{aff44}
\and
Institute for Astronomy, University of Edinburgh, Royal Observatory, Blackford Hill, Edinburgh EH9 3HJ, UK\label{aff45}
\and
Jodrell Bank Centre for Astrophysics, Department of Physics and Astronomy, University of Manchester, Oxford Road, Manchester M13 9PL, UK\label{aff46}
\and
European Space Agency/ESRIN, Largo Galileo Galilei 1, 00044 Frascati, Roma, Italy\label{aff47}
\and
Universit\'e Lyon 1, CNRS/IN2P3, IP2I Lyon, Villeurbanne, France\label{aff48}
\and
Institut de Ci\`{e}ncies del Cosmos (ICCUB), Universitat de Barcelona (IEEC-UB), Mart\'{i} i Franqu\`{e}s 1, 08028 Barcelona, Spain\label{aff49}
\and
Instituci\'o Catalana de Recerca i Estudis Avan\c{c}ats (ICREA), Passeig de Llu\'{\i}s Companys 23, 08010 Barcelona, Spain\label{aff50}
\and
Institut de Ciencies de l'Espai (IEEC-CSIC), Campus UAB, Carrer de Can Magrans, s/n Cerdanyola del Vall\'es, 08193 Barcelona, Spain\label{aff51}
\and
Universit\'e Lyon 1, CNRS/IN2P3, IUF, IP2I Lyon, Villeurbanne, France \label{aff52}
\and
Canada-France-Hawaii Telescope, 65-1238 Mamalahoa Hwy, Kamuela, HI 96743, USA\label{aff53}
\and
Department of Astronomy, University of Geneva, ch. d'Ecogia 16, 1290 Versoix, Switzerland\label{aff54}
\and
Universit\'e Paris-Saclay, CNRS, Institut d'astrophysique spatiale, 91405, Orsay, France\label{aff55}
\and
INFN-Padova, Via Marzolo 8, 35131 Padova, Italy\label{aff56}
\and
Aix-Marseille Universit\'e, CNRS/IN2P3, CPPM, Marseille, France\label{aff57}
\and
INAF-Istituto di Astrofisica e Planetologia Spaziali, via del Fosso del Cavaliere, 100, 00100 Roma, Italy\label{aff58}
\and
INFN-Bologna, Via Irnerio 46, 40126 Bologna, Italy\label{aff59}
\and
Department of Physics, University of Oxford, Keble Road, Oxford OX1 3RH, UK\label{aff60}
\and
University Observatory, LMU Faculty of Physics, Scheinerstr.~1, 81679 Munich, Germany\label{aff61}
\and
INFN-Sezione di Milano, Via Celoria 16, 20133 Milano, Italy\label{aff62}
\and
National Research Council, Herzberg Astronomy and Astrophysics Research Centre, 5071 W. Saanich Rd. Victoria, BC, V9E 2E7, Canada\label{aff63}
\and
Institute of Theoretical Astrophysics, University of Oslo, P.O. Box 1029 Blindern, 0315 Oslo, Norway\label{aff64}
\and
California Institute of Technology, 1200 E California Blvd, Pasadena, CA 91125, USA\label{aff65}
\and
Caltech/IPAC, 1200 E. California Blvd., Pasadena, CA 91125, USA\label{aff66}
\and
Leiden Observatory, Leiden University, Einsteinweg 55, 2333 CC Leiden, The Netherlands\label{aff67}
\and
Jet Propulsion Laboratory, California Institute of Technology, 4800 Oak Grove Drive, Pasadena, CA, 91109, USA\label{aff68}
\and
Felix Hormuth Engineering, Goethestr. 17, 69181 Leimen, Germany\label{aff69}
\and
Technical University of Denmark, Elektrovej 327, 2800 Kgs. Lyngby, Denmark\label{aff70}
\and
Cosmic Dawn Center (DAWN), Denmark\label{aff71}
\and
Max-Planck-Institut f\"ur Astronomie, K\"onigstuhl 17, 69117 Heidelberg, Germany\label{aff72}
\and
NASA Goddard Space Flight Center, Greenbelt, MD 20771, USA\label{aff73}
\and
Universit\'e de Gen\`eve, D\'epartement de Physique Th\'eorique and Centre for Astroparticle Physics, 24 quai Ernest-Ansermet, CH-1211 Gen\`eve 4, Switzerland\label{aff74}
\and
Helsinki Institute of Physics, Gustaf H{\"a}llstr{\"o}min katu 2, University of Helsinki, 00014 Helsinki, Finland\label{aff75}
\and
Laboratoire d'etude de l'Univers et des phenomenes eXtremes, Observatoire de Paris, Universit\'e PSL, Sorbonne Universit\'e, CNRS, 92190 Meudon, France\label{aff76}
\and
SKAO, Jodrell Bank, Lower Withington, Macclesfield SK11 9FT, UK\label{aff77}
\and
Centre de Calcul de l'IN2P3/CNRS, 21 avenue Pierre de Coubertin 69627 Villeurbanne Cedex, France\label{aff78}
\and
University of Applied Sciences and Arts of Northwestern Switzerland, School of Computer Science, 5210 Windisch, Switzerland\label{aff79}
\and
Universit\"at Bonn, Argelander-Institut f\"ur Astronomie, Auf dem H\"ugel 71, 53121 Bonn, Germany\label{aff80}
\and
INFN-Sezione di Roma, Piazzale Aldo Moro, 2 - c/o Dipartimento di Fisica, Edificio G. Marconi, 00185 Roma, Italy\label{aff81}
\and
Department of Physics, Institute for Computational Cosmology, Durham University, South Road, Durham, DH1 3LE, UK\label{aff82}
\and
Universit\'e Paris Cit\'e, CNRS, Astroparticule et Cosmologie, 75013 Paris, France\label{aff83}
\and
CNRS-UCB International Research Laboratory, Centre Pierre Bin\'etruy, IRL2007, CPB-IN2P3, Berkeley, USA\label{aff84}
\and
Institute of Physics, Laboratory of Astrophysics, Ecole Polytechnique F\'ed\'erale de Lausanne (EPFL), Observatoire de Sauverny, 1290 Versoix, Switzerland\label{aff85}
\and
Dipartimento di Fisica - Sezione di Astronomia, Universit\`a di Trieste, Via Tiepolo 11, 34131 Trieste, Italy\label{aff86}
\and
Telespazio UK S.L. for European Space Agency (ESA), Camino bajo del Castillo, s/n, Urbanizacion Villafranca del Castillo, Villanueva de la Ca\~nada, 28692 Madrid, Spain\label{aff87}
\and
Institut de F\'{i}sica d'Altes Energies (IFAE), The Barcelona Institute of Science and Technology, Campus UAB, 08193 Bellaterra (Barcelona), Spain\label{aff88}
\and
School of Mathematics and Physics, University of Surrey, Guildford, Surrey, GU2 7XH, UK\label{aff89}
\and
European Space Agency/ESTEC, Keplerlaan 1, 2201 AZ Noordwijk, The Netherlands\label{aff90}
\and
Waterloo Centre for Astrophysics, University of Waterloo, Waterloo, Ontario N2L 3G1, Canada\label{aff91}
\and
Department of Physics and Astronomy, University of Waterloo, Waterloo, Ontario N2L 3G1, Canada\label{aff92}
\and
Perimeter Institute for Theoretical Physics, Waterloo, Ontario N2L 2Y5, Canada\label{aff93}
\and
Universit\'e Paris-Saclay, Universit\'e Paris Cit\'e, CEA, CNRS, AIM, 91191, Gif-sur-Yvette, France\label{aff94}
\and
Space Science Data Center, Italian Space Agency, via del Politecnico snc, 00133 Roma, Italy\label{aff95}
\and
Centre National d'Etudes Spatiales -- Centre spatial de Toulouse, 18 avenue Edouard Belin, 31401 Toulouse Cedex 9, France\label{aff96}
\and
Institute of Space Science, Str. Atomistilor, nr. 409 M\u{a}gurele, Ilfov, 077125, Romania\label{aff97}
\and
Consejo Superior de Investigaciones Cientificas, Calle Serrano 117, 28006 Madrid, Spain\label{aff98}
\and
Universidad de La Laguna, Dpto. Astrof\'\i sica, E-38206 La Laguna, Tenerife, Spain\label{aff99}
\and
Dipartimento di Fisica e Astronomia "G. Galilei", Universit\`a di Padova, Via Marzolo 8, 35131 Padova, Italy\label{aff100}
\and
Instituto de F\'isica Te\'orica UAM-CSIC, Campus de Cantoblanco, 28049 Madrid, Spain\label{aff101}
\and
Institut de Recherche en Astrophysique et Plan\'etologie (IRAP), Universit\'e de Toulouse, CNRS, UPS, CNES, 14 Av. Edouard Belin, 31400 Toulouse, France\label{aff102}
\and
Universit\'e St Joseph; Faculty of Sciences, Beirut, Lebanon\label{aff103}
\and
Departamento de F\'isica, FCFM, Universidad de Chile, Blanco Encalada 2008, Santiago, Chile\label{aff104}
\and
Universit\"at Innsbruck, Institut f\"ur Astro- und Teilchenphysik, Technikerstr. 25/8, 6020 Innsbruck, Austria\label{aff105}
\and
Department of Physics and Helsinki Institute of Physics, Gustaf H\"allstr\"omin katu 2, University of Helsinki, 00014 Helsinki, Finland\label{aff106}
\and
Departamento de F\'isica, Faculdade de Ci\^encias, Universidade de Lisboa, Edif\'icio C8, Campo Grande, PT1749-016 Lisboa, Portugal\label{aff107}
\and
Instituto de Astrof\'isica e Ci\^encias do Espa\c{c}o, Faculdade de Ci\^encias, Universidade de Lisboa, Tapada da Ajuda, 1349-018 Lisboa, Portugal\label{aff108}
\and
Mullard Space Science Laboratory, University College London, Holmbury St Mary, Dorking, Surrey RH5 6NT, UK\label{aff109}
\and
Cosmic Dawn Center (DAWN)\label{aff110}
\and
Niels Bohr Institute, University of Copenhagen, Jagtvej 128, 2200 Copenhagen, Denmark\label{aff111}
\and
Universidad Polit\'ecnica de Cartagena, Departamento de Electr\'onica y Tecnolog\'ia de Computadoras,  Plaza del Hospital 1, 30202 Cartagena, Spain\label{aff112}
\and
European University of Technology EUt+, European Union\label{aff113}
\and
Centre for Information Technology, University of Groningen, P.O. Box 11044, 9700 CA Groningen, The Netherlands\label{aff114}
\and
Institute of Space Sciences (ICE, CSIC), Campus UAB, Carrer de Can Magrans, s/n, 08193 Barcelona, Spain\label{aff115}
\and
Institut d'Estudis Espacials de Catalunya (IEEC),  Edifici RDIT, Campus UPC, 08860 Castelldefels, Barcelona, Spain\label{aff116}
\and
Istituto Nazionale di Fisica Nucleare, Sezione di Ferrara, Via Giuseppe Saragat 1, 44122 Ferrara, Italy\label{aff117}
\and
ICSC - Centro Nazionale di Ricerca in High Performance Computing, Big Data e Quantum Computing, Via Magnanelli 2, Bologna, Italy\label{aff118}
\and
INAF, Istituto di Radioastronomia, Via Piero Gobetti 101, 40129 Bologna, Italy\label{aff119}
\and
Universit\'e C\^{o}te d'Azur, Observatoire de la C\^{o}te d'Azur, CNRS, Laboratoire Lagrange, Bd de l'Observatoire, CS 34229, 06304 Nice cedex 4, France\label{aff120}
\and
Department of Mathematical Physics, Institute of Physics, University of S\~ao Paulo, R. do Mat\~ao 1371, 05508-090, S\~ao Paulo, SP, Brazil\label{aff121}
\and
Oskar Klein Centre for Cosmoparticle Physics, Department of Physics, Stockholm University, Stockholm, SE-106 91, Sweden\label{aff122}
\and
Astrophysics Group, Blackett Laboratory, Imperial College London, London SW7 2AZ, UK\label{aff123}
\and
Univ. Grenoble Alpes, CNRS, Grenoble INP, LPSC-IN2P3, 53, Avenue des Martyrs, 38000 Grenoble, France\label{aff124}
\and
Dipartimento di Fisica, Sapienza Universit\`a di Roma, Piazzale Aldo Moro 2, 00185 Roma, Italy\label{aff125}
\and
Department of Mathematics and Physics E. De Giorgi, University of Salento, Via per Arnesano, CP-I93, 73100, Lecce, Italy\label{aff126}
\and
INFN, Sezione di Lecce, Via per Arnesano, CP-193, 73100, Lecce, Italy\label{aff127}
\and
INAF-Sezione di Lecce, c/o Dipartimento Matematica e Fisica, Via per Arnesano, 73100, Lecce, Italy\label{aff128}
\and
Institut d'Astrophysique de Paris, 98bis Boulevard Arago, 75014, Paris, France\label{aff129}
\and
ICL, Junia, Universit\'e Catholique de Lille, LITL, 59000 Lille, France\label{aff130}
\and
CERCA/ISO, Department of Physics, Case Western Reserve University, 10900 Euclid Avenue, Cleveland, OH 44106, USA\label{aff131}
\and
Laboratoire Univers et Th\'eorie, Observatoire de Paris, Universit\'e PSL, Universit\'e Paris Cit\'e, CNRS, 92190 Meudon, France\label{aff132}
\and
Departamento de F{\'\i}sica Fundamental. Universidad de Salamanca. Plaza de la Merced s/n. 37008 Salamanca, Spain\label{aff133}
\and
Universiteit Gent, Department of Physics and Astronomy, Proeftuinstraat 86 N3, 9000 Ghent, Belgium
\label{aff134}
\and
Universit\'e de Strasbourg, CNRS, Observatoire astronomique de Strasbourg, UMR 7550, 67000 Strasbourg, France\label{aff135}
\and
Center for Data-Driven Discovery, Kavli IPMU (WPI), UTIAS, The University of Tokyo, Kashiwa, Chiba 277-8583, Japan\label{aff136}
\and
Max-Planck-Institut f\"ur Physik, Boltzmannstr. 8, 85748 Garching, Germany\label{aff137}
\and
Centre de Recherche Astrophysique de Lyon, UMR5574, CNRS, Universit\'e Claude Bernard Lyon 1, ENS de Lyon, 69230, Saint-Genis-Laval, France\label{aff138}
\and
Department of Physics \& Astronomy, University of California Irvine, Irvine CA 92697, USA\label{aff139}
\and
Kapteyn Astronomical Institute, University of Groningen, PO Box 800, 9700 AV Groningen, The Netherlands\label{aff140}
\and
Departamento F\'isica Aplicada, Universidad Polit\'ecnica de Cartagena, Campus Muralla del Mar, 30202 Cartagena, Murcia, Spain\label{aff141}
\and
Department of Computer Science, Aalto University, PO Box 15400, Espoo, FI-00 076, Finland\label{aff142}
\and
School of Physics and Astronomy, University of Nottingham, University Park, Nottingham NG7 2RD, UK\label{aff143}
\and
Ruhr University Bochum, Faculty of Physics and Astronomy, Astronomical Institute (AIRUB), German Centre for Cosmological Lensing (GCCL), 44780 Bochum, Germany\label{aff144}
\and
DARK, Niels Bohr Institute, University of Copenhagen, Jagtvej 155, 2200 Copenhagen, Denmark\label{aff145}
\and
Department of Physics, Centre for Extragalactic Astronomy, Durham University, South Road, Durham, DH1 3LE, UK\label{aff146}
\and
Astrophysics Research Centre, University of KwaZulu-Natal, Westville Campus, Durban 4041, South Africa\label{aff147}
\and
School of Mathematics, Statistics \& Computer Science, University of KwaZulu-Natal, Westville Campus, Durban 4041, South Africa\label{aff148}
\and
Department of Physics and Astronomy, Vesilinnantie 5, University of Turku, 20014 Turku, Finland\label{aff149}
\and
Serco for European Space Agency (ESA), Camino bajo del Castillo, s/n, Urbanizacion Villafranca del Castillo, Villanueva de la Ca\~nada, 28692 Madrid, Spain\label{aff150}
\and
ARC Centre of Excellence for Dark Matter Particle Physics, Melbourne, Australia\label{aff151}
\and
Centre for Astrophysics \& Supercomputing, Swinburne University of Technology,  Hawthorn, Victoria 3122, Australia\label{aff152}
\and
Department of Physics and Astronomy, University of the Western Cape, Bellville, Cape Town, 7535, South Africa\label{aff153}
\and
MTA-CSFK Lend\"ulet Large-Scale Structure Research Group, Konkoly-Thege Mikl\'os \'ut 15-17, H-1121 Budapest, Hungary\label{aff154}
\and
Konkoly Observatory, HUN-REN CSFK, MTA Centre of Excellence, Budapest, Konkoly Thege Mikl\'os {\'u}t 15-17. H-1121, Hungary\label{aff155}
\and
Institute for Theoretical Particle Physics and Cosmology (TTK), RWTH Aachen University, 52056 Aachen, Germany\label{aff156}
\and
INAF-Osservatorio Astrofisico di Arcetri, Largo E. Fermi 5, 50125, Firenze, Italy\label{aff157}
\and
Centro de Astrof\'{\i}sica da Universidade do Porto, Rua das Estrelas, 4150-762 Porto, Portugal\label{aff158}
\and
HE Space for European Space Agency (ESA), Camino bajo del Castillo, s/n, Urbanizacion Villafranca del Castillo, Villanueva de la Ca\~nada, 28692 Madrid, Spain\label{aff159}
\and
STAR Institute, University of Li{\`e}ge, Quartier Agora, All\'ee du six Ao\^ut 19c, 4000 Li\`ege, Belgium\label{aff160}
\and
Instituto de Astrof\'isica e Ci\^encias do Espa\c{c}o, Faculdade de Ci\^encias, Universidade de Lisboa, Campo Grande, 1749-016 Lisboa, Portugal\label{aff161}
\and
Department of Astrophysics, University of Zurich, Winterthurerstrasse 190, 8057 Zurich, Switzerland\label{aff162}
\and
Department of Mathematics and Physics, Roma Tre University, Via della Vasca Navale 84, 00146 Rome, Italy\label{aff163}
\and
Theoretical astrophysics, Department of Physics and Astronomy, Uppsala University, Box 524, 751 20 Uppsala, Sweden\label{aff164}
\and
Mathematical Institute, University of Leiden, Einsteinweg 55, 2333 CA Leiden, The Netherlands\label{aff165}
\and
School of Physics \& Astronomy, University of Southampton, Highfield Campus, Southampton SO17 1BJ, UK\label{aff166}
\and
Institute of Astronomy, University of Cambridge, Madingley Road, Cambridge CB3 0HA, UK\label{aff167}
\and
Univ. Lille, CNRS, Centrale Lille, UMR 9189 CRIStAL, 59000 Lille, France\label{aff168}
\and
Department of Physics and Astronomy, University of California, Davis, CA 95616, USA\label{aff169}
\and
Center for Astrophysics and Cosmology, University of Nova Gorica, Nova Gorica, Slovenia\label{aff170}
\and
Department of Astrophysical Sciences, Peyton Hall, Princeton University, Princeton, NJ 08544, USA\label{aff171}
\and
Fakult\"at f\"ur Physik, Universit\"at Bielefeld, Postfach 100131, 33501 Bielefeld, Germany\label{aff172}
\and
Space physics and astronomy research unit, University of Oulu, Pentti Kaiteran katu 1, FI-90014 Oulu, Finland\label{aff173}
\and
Department of Physics and Astronomy, Lehman College of the CUNY, Bronx, NY 10468, USA\label{aff174}
\and
American Museum of Natural History, Department of Astrophysics, New York, NY 10024, USA\label{aff175}
\and
SRON Netherlands Institute for Space Research, Landleven 12, 9747 AD, Groningen, The Netherlands\label{aff176}
\and
Department of Physics and Astronomy, University of British Columbia, Vancouver, BC V6T 1Z1, Canada\label{aff177}}    

 \abstract{
Applications of strong gravitational lensing in astrophysics and 
cosmology critically depend on building reliable strong lensing models, which require accurate measurements of on-sky positions of multiple images and of the redshifts of the background sources from which they originate. In this context, \Euclid is a unique facility, as it simultaneously provides high-resolution multi-band photometric imaging and slitless spectroscopy for an unprecedentedly large sample of gravitational lenses, most of which are newly discovered. In this work, we exploit \Euclid slitless spectroscopic data to determine the redshifts of four spatially extended lensed sources, and to reconstruct emission-line images.  Specifically, two of the target sources are lensed into giant arcs by the massive galaxy clusters SPT-CLJ0411$-$4819 ($z_\mathrm{d}=0.424$, where ``d'' denotes the lensing deflector) and Abell 2280 ($z_\mathrm{d}=0.326$). The third system is a quadruply imaged active galactic nucleus (AGN) candidate strongly lensed by a massive galaxy at $z_\mathrm{d}=0.68$, while the fourth is a galaxy-scale Einstein ring produced by a lens galaxy at $z_\mathrm{d}=0.235$. This work represents a proof of concept for the exploitation of \Euclid slitless spectroscopic data in the study of strong gravitational lensing, providing one of the first determinations of spectroscopic redshifts for strongly lensed features from NISP spectra. These results demonstrate the feasibility of applying the same methodology to the very large sample of gravitational lenses that \Euclid has started to discover and is expected to identify over the course of the mission.}
 
%
%
%
%
   \titlerunning{NISP spectroscopy of gravitational arcs and multiple images.}
   \authorrunning{Euclid Collaboration: P. Bergamini et al.}
   
   \maketitle
%
   
\section{\label{sc:Intro}Introduction}

Strong gravitational lensing is one of the most powerful techniques for probing the inner total mass distribution of galaxies and galaxy clusters \citep[e.g.,][]{Lagattuta2026,Rihtarsic2026,Beauchesne2026,Cerny2026,Acebron_2025,Vegetti_2024,Patel2024,Bergamini2023,Diego_2023_SLGC,Caminha_2023,Mahler2023,Allingham_2023,Jauzac2021,Meneghetti_2020,Zitrin2020,Richard2015}, as well as for studying the physical properties of magnified (up to several orders of magnitude) high-redshift galaxies \citep[e.g.,][]{Newman2026,Vanzella_2026,Messa_2026,Mestric2022, Bradac_2025,Adamo2024,Castellano2023}. Moreover, strong lensing can provide competitive estimates of cosmological parameter values (e.g., $H_0$, $\Omega_{\rm m}$, $\Omega_{\Lambda}$) by exploiting the ratio of the angular diameter distances of multiply-lensed sources at different redshifts \citep[][]{Caminha2022,Magana2018,Acebron2017,Jullo2010} and the measured time delays between multiple images of time-variable background sources \citep[][]{Suyu_2026,Grillo2024}. 

The robustness of strong lensing models critically depends on precise measurements of the positions of the largest possible sample of multiple images associated with distinct background sources, which provide the primary constraints on the lens mass distribution, as well as on accurate spectroscopic determinations of the lens and source redshifts, from which the relevant cosmological distances can be inferred. In this context, the spectro-photometric data provided by the \Euclid space telescope represent an unprecedented dataset for strong lensing studies \citep[e.g.][]{DR1-SL01,Q1-SP048,Q1-SP057}. 

Upon completion, the Euclid Wide Survey \citep[EWS,][]{Scaramella_2022} will cover approximately 14\,000 deg$^2$ of the sky. It will be complemented by the Euclid Deep Survey (EDS; \citealt{EuclidSkyOverview}), which will provide \Euclid spectro-photometric data over an area of 53 deg$^2$, reaching approximately two magnitudes deeper than the EWS in the \HE band \citep{EuclidSkyNISP,Schirmer-EP18}. 
Together, these surveys will dramatically increase the number of known gravitational lensing systems, the majority of which will lack spectroscopic coverage from complementary facilities. In general, the crowded environments of galaxy cluster fields, together with the small angular separations and complex morphologies characteristic of galaxy-scale strong lensing systems, make the automated redshift determination with the \Euclid pipeline extremely challenging.

In this work, we exploit the advanced {\tt Python}-based data analysis tool {\tt GELSA} \citep{DR1-Gabarra} to perform a customised extraction of the spectra of four strongly lensed sources from \Euclid Near-Infrared Spectrometer and Photometer \citep[NISP,][]{EuclidSkyNISP} slitless spectroscopic data. Through the identification of prominent emission lines in the extracted spectra, we obtain robust spectroscopic redshift measurements for the lensed background sources.

Two of these sources are tangential giant arcs produced by the massive galaxy clusters SPT-CLJ0411$-$4819 (\Fig\ref{fig:SPT-CLJ0411-4819}) and Abell~2280 (\Fig\ref{fig:Abell2280}). Both arcs were detected in the Euclid Quick Release Q1 dataset, which covers an area of 63.1\,$\mathrm{deg}^2$, by \citeauthor{Q1-SP057} (\citeyear{Q1-SP057}; B26 hereafter). The third system, [WH2024] J181607.9$+$681651 (hereafter WHJ181607.9$+$681651, see \Fig\ref{fig:J181607}), independently identified in the Q1 dataset by B26 and \cite{Q1-SP048}, consists of a candidate active galactic nucleus (AGN) lensed into an Einstein-cross configuration (see also \citealt{Q1-SP053} and \citealt{Q1-Vincken}). Finally, PS1J1821+7130 (\Fig\ref{fig:PS1J1821+7130}), recently identified in the \cite{DR1cite} observations \citep{DR1-TP001}, exhibits a nearly complete Einstein ring with several compact substructures \citep[][Euclid Collaboration: Vincken et al., in prep.; Euclid Collaboration: Fogliardi et al., in prep.]{DR1-SL01}.

The four systems were selected as representative examples of cluster- and galaxy-scale strong gravitational lenses already discovered by \Euclid. However, we emphasise that this study serves as a pilot proof of concept, demonstrating the feasibility of applying the approach presented here to a much larger sample of gravitational lenses discovered by \Euclid, a substantial fraction of which are expected to exhibit prominent emission-line features in their NISP spectra.

The paper is organised as follows. In \Sec\ref{sc:SelectedStrongLensingSystems}, we describe the four strong lensing systems analysed in this work. In \Sec\ref{sc:Data}, we briefly present the spectrophotometric data used in this work. The {\tt Python} codes and the procedure adopted for spectral extraction are described in \Sec\ref{sc:SpectralExtraction}. The results of our study are presented in \Sec\ref{sc:Results}, while in \Sec\ref{sc:Conclusions} we draw our conclusions. Throughout the paper, we assume a flat ${\Lambda\mathrm{CDM}}$ cosmology with $H_0 = 70$ $ \mathrm{km~s^{-1}~Mpc^{-1}}$, and matter density $\Omega_{\rm m}=0.3$. Magnitudes are given in the AB system \citep{Oke1974}.

\section{\label{sc:SelectedStrongLensingSystems}Selected strong lensing systems}

Three of the four analysed strong lensing systems, namely SPT-CLJ0411$-$4819, Abell~2280, and WHJ181607.9+681651 (Table\,\ref{tab:table_redshifts} and Figs.\,\ref{fig:SPT-CLJ0411-4819}, \ref{fig:Abell2280}, and \ref{fig:J181607}), were classified as secure strong gravitational lenses by B26. These systems exhibit bright strong lensing features, with integrated magnitudes of $\HE<21.5$. Specifically, SPT-CLJ0411$-$4819 and Abell~2280 are cluster-scale lenses located in the Euclid Deep Field South (EDF-S) and North (EDF-N), respectively, whereas WHJ181607.9$+$681651 is a galaxy-scale lens in EDF-N (see also \citealt{Q1-SP048}). The remaining system, PS1J1821+7130 (\Fig\ref{fig:PS1J1821+7130}), is a galaxy-galaxy strong lensing system identified in the northern Euclid DR1 area using neural networks specifically designed to find strong lenses in \Euclid imaging data \citep[][Euclid Collaboration: Vincken et al., in prep.; Euclid Collaboration: Fogliardi et al., in prep.]{DR1-SL01}. Below, we briefly describe each of the four systems.

\subsection{\label{sc:SPT-CLJ0411-4819}SPT-CLJ0411\texorpdfstring{$-$}{-}4819 (aka ACT-CLJ0411.2\texorpdfstring{$-$}{-}4819)}

The galaxy cluster (\Fig\,\ref{fig:SPT-CLJ0411-4819}) was identified through the Sunyaev--Zeldovich effect by \cite{Planck_2011} and independently detected in the South Pole Telescope \citep[SPT-SZ,][]{Bleem_2015,Williamson_2011} and Atacama Cosmology Telescope \citep[ACT-SZ,][]{Hilton_2021} surveys. In particular, SPT-SZ and ACT-SZ measured the cluster mass, $M_{500{\rm c}}$, to be $\sim8.2\times10^{14}$\SI{}{\solarmass} and $\sim5.8\times10^{14}$\SI{}{\solarmass}, respectively.

Spectroscopic observations of the cluster were obtained as part of the SPT-GMOS programme \citep[][ID: GS-2012B-Q-59]{Bayliss_2016} using the Gemini Multi-Object Spectrograph (GMOS) on Gemini South. \cite{Bayliss_2016} confirmed 44 cluster member galaxies. From these data, they obtained a median cluster redshift of $0.4241\pm0.006$ and a cluster velocity dispersion of $(1267\pm177)\,$\SI{}{\kms}. 

The cluster was subsequently identified by \cite{Wen_2024} as an over-density in the stellar mass distribution of galaxies at a similar redshift (which corresponds to [WH2024]  J041111.0$-$481940 in their catalogue), using multi-band data from the Dark Energy Spectroscopic Instrument (DESI) Legacy Imaging Surveys \citep[Data Release 9 and 10;][]{Dey_2019,Zou_2017} and the
Wide-field Infrared Survey Explorer \citep[WISE,][]{Wright_2010}. \cite{Wen_2024} identified 113 photometric cluster members and measured a cluster mass of $M_{500{\rm c}}$ $\sim5.3\times10^{14}$\SI{}{\solarmass} based on the estimated cluster richness.

The giant gravitational arc analysed in this work (\Fig\,\ref{fig:SPT-CLJ0411-4819}) was first identified by \cite{Bayliss_2016}. However, the low signal-to-noise ratio (${\rm S/N}$) of the SPT-GMOS spectrum precluded a secure redshift determination for the lensed source. The cluster was also recognised as a strong gravitational lens by convolutional neural networks (CNNs) developed by \cite{Jacobs_2019} and applied to the Dark Energy Survey \citep[DES;][]{DES_2005, DES_2016} imaging data to search for galaxy-scale gravitational lenses. Finally, the strong lensing nature of the system was unequivocally confirmed by B26, who identified several additional strong lensing features in the cluster field thanks to Euclid Q1 imaging data.

\subsection{\label{sc:Abell2280}Abell~2280}
Spectroscopic observations of Abell~2280 were first carried out with the Imaging Grism Instrument, mounted on the McDonald Observatory 107-inch Telescope \citep[University of Texas,][]{Lacy_1993}, yielding a redshift estimate of $z=0.33$ for the brightest cluster galaxy (BCG). This value was subsequently refined using multi-object spectroscopic data for seven cluster galaxies obtained at the Canada-France-Hawaii Telescope \citep[CFHT,][]{Gioia_1995}, yielding a median cluster redshift of $z_\mathrm{d}^\mathrm{sp}=0.326\pm0.002$. The latter study also reported the first detection of the giant red arc analysed in this work (encircled in green in \Fig\ref{fig:Abell2280}), with a $R$-band magnitude of $R=22.9\pm0.2$ and no detection in the $B$ band.
More recently, \cite{Wen_2024} photometrically identified 73 galaxies as members of Abell~2280 (which corresponds to [WH2024] J174330.4$+$634142 in their catalogue) and measured a cluster mass of $M_{500{\rm c}}\sim4.7\times10^{14}$\SI{}{\solarmass}.

\subsection{WHJ181607.9+681651}
\label{sc:WHJ181607.9+681651}
This system was identified by \cite{Wen_2024} as a candidate group of 14 galaxies at a photometric redshift of $z^{\rm ph}\sim0.674$. In particular, they measured a photometric redshift of $z_\mathrm{d}^\mathrm{ph}=0.68\pm0.04$ for the early-type elliptical galaxy acting as a gravitational lens, located at the centre of the RGB cutout shown in \Fig\ref{fig:J181607}. Subsequent visual inspection of Q1 data by B26 revealed the galaxy-scale strong lensing nature of the system, which was assigned the maximum lensing grade ($\mathcal{P}_\mathrm{lens}=1$, see B26). As shown in the bottom-left panel of \Fig\ref{fig:J181607}, the lens galaxy produces four images of a red background source arranged in an Einstein cross configuration. The nucleated morphology of these lensed images suggests that the source may consist of an AGN and its host galaxy.

\subsection{PS1J1821+7130}
The system was initially discovered by \cite{Canameras_2020} as part of the Highly Optimized Lensing Investigations of Supernovae, Microlensing Objects, and Kinematics of Ellipticals and Spirals (HOLISMOKES) collaboration \citep{Suyu_2020}. The discovery was made using a CNN specifically trained to identify galaxy-scale strong lenses across the $30\,000$ deg$^2$ Pan-STARRS 3$\pi$ survey area in the northern sky. PS1J1821+7130 was assigned the highest possible confidence of being a real lens, with a CNN lensing probability equal to unity. 

The system was subsequently rediscovered by \cite{DR1-SL01} using machine-learning algorithms designed to identify galaxy-scale strong lensing systems in Euclid DR1 observations. The \Euclid images unequivocally reveal the strong lensing nature of PS1J1821+7130, showing a nearly complete Einstein ring populated by several multiply imaged substructures (see the bottom-left panel of \Fig\,\ref{fig:PS1J1821+7130}). A spectroscopic redshift of $z_\mathrm{d}^\mathrm{sp}=0.235\pm0.001$ is reported for the deflector galaxy in DESI DR1 dataset \citep{DESI_2026,DESI_2024}.

\section{\label{sc:Data}\Euclid spectrophotometric dataset}

The results presented in this work are primarily based on spectroscopic data obtained with the \Euclid NISP slitless spectroscopic channel \citep[NISP-S,][]{EuclidSkyNISP}.
These data were used to identify the emission-line features employed to determine the redshifts of the lensed sources with the methodology described in \Sec\ref{sc:SpectralExtraction}.
Specifically, the NISP spectra were obtained with the two red grisms, RGS000 and RGS180, at all orientations available as part of DR1. For PS1J1821+7130, which lies within the EWS area, these correspond to RGS000+0, RGS000$-$4, RGS180+0, and RGS180+4 \citep[][]{Q1-TP006}, with the latter three rotated by $-$4, 180, and 184 degrees, respectively, relative to RGS000+0. In contrast, SPT-CLJ0411$-$4819, Abell~2280, and WHJ181607.9$+$681651 have so far been observed with 38, 9, and 3 \Euclid spectroscopic exposures, respectively, corresponding to a comparable number of distinct dispersion orientations, as part of the EDS (Table\,\ref{tab:table_redshifts}). Each spectroscopic exposure has an integration time of 550 seconds. We note that the availability of multiple orientations is important not only for increasing the ${\rm S/N}$ of the resulting stacked spectrum but also for mitigating spectral contamination from nearby bright sources. Both red grisms have a resolving power of $R>480$ \citep{Q1-TP006} and produce spectra covering a wavelength range of approximately $1206\text{--}1892\,$nm with a dispersion of $1.372\,$nm\,px$^{-1}$ \citep{DR1-TP024}. At the EWS depth, NISP-S is capable of detecting emission lines down to a line flux of approximately $2 \times 10^{-16}\,\mathrm{erg\, cm^{-2}\, s^{-1}}$ at a significance level of $3.5\,\sigma$, as estimated at a wavelength of 1.6$\,$\textmu m for a source with a diameter of \ang{;;0.5}. This line-flux sensitivity is expected to improve further, reaching $5 \times 10^{-17}\,\mathrm{erg\, cm^{-2}\, s^{-1}}$ at the full depth of the EDS \citep{EuclidSkyOverview}. Although blue-grism observations will also be conducted in the EDS, the present analysis relies on all currently available red-grism data, corresponding to the standard configuration adopted in the EWS.

\begin{figure*}
\centering
\includegraphics[angle=0,width=1.0\linewidth]{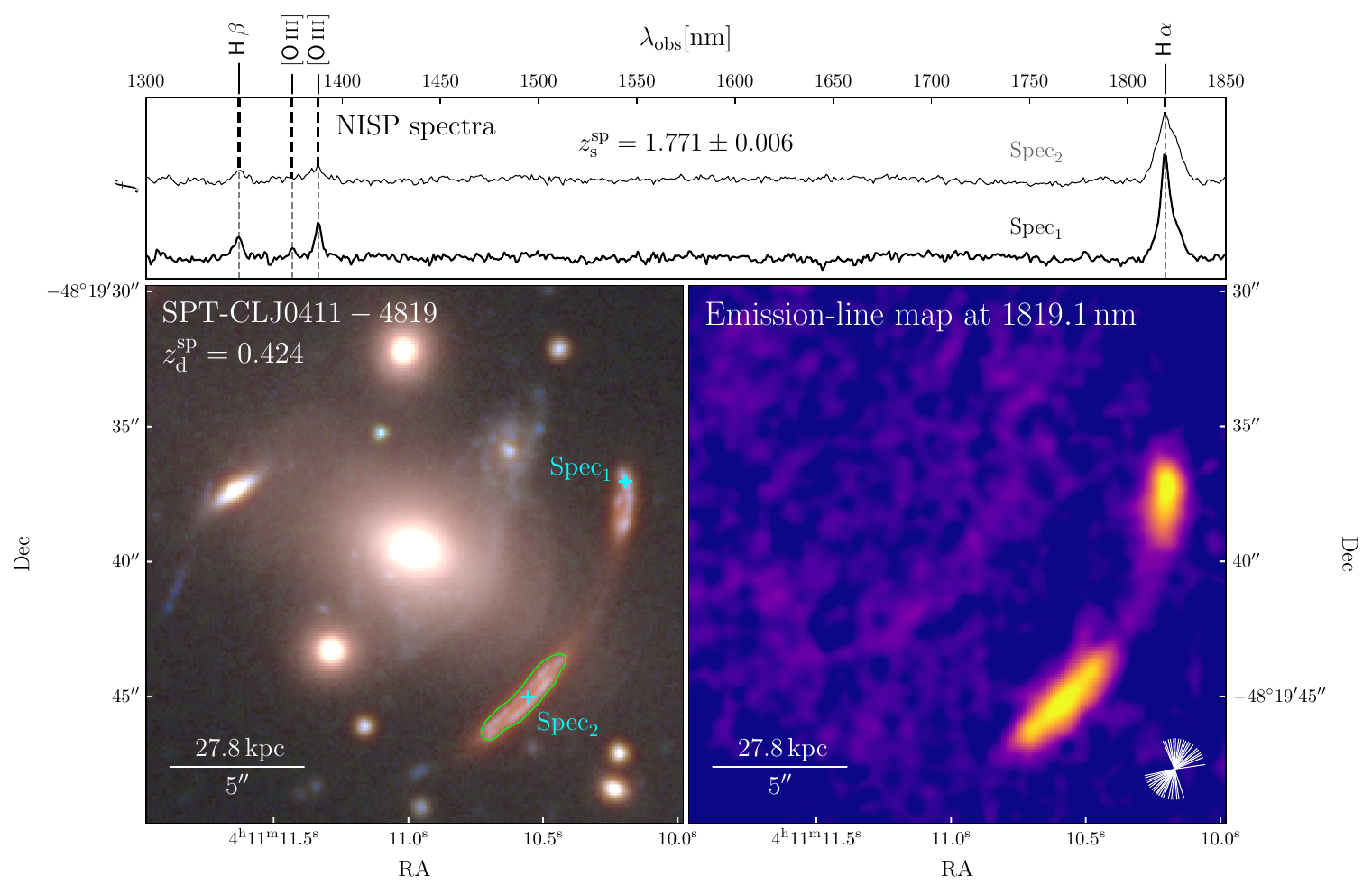}
\caption{\Euclid data of the galaxy cluster SPT-CLJ0411$-$4819 at $z_\mathrm{d}^\mathrm{sp}=0.424\pm0.006$, which lenses a background source with a measured \Euclid spectroscopic redshift of $z_\mathrm{s}^\mathrm{sp}=1.771\pm0.006$ into a giant gravitational arc. {\it Bottom-left:} RGB image of the cluster obtained by combining \Euclid photometric bands (\HE, \YE, \JE, and \IE) using the \texttt{azulero} code \citep{Basset_Azulero}. The spatial scale corresponding to \ang{;;5} refers to the deflector redshift. The green polygon corresponds to the aperture used to measure the magnitude of the gravitational arc in different bands and the associated uncertainty, as described in Appendix\,\ref{sc:PhotometricRedshift}. {\it Bottom-right:} Emission-line map at 1819.1\,nm, showing the \ha emission associated with the arc. The map is obtained by combining 38 \Euclid dispersed spectroscopic images; the corresponding dispersion directions are indicated by the white segments in the bottom-right corner of the panel. The map, oversampled at the native pixel scale of the \IE filter (\ang{;;0.1}), is smoothed with a Gaussian kernel with a standard deviation of two pixels (\ang{;;0.2}). {\it Top:} Stacked \Euclid NISP-S spectra (in arbitrary units) extracted at two positions, ``$\mathrm{Spec_1}$'' and ``$\mathrm{Spec_2}$’’, marked by cyan crosses in the RGB image. The spectral extraction is performed over a spatial aperture of three NISP pixels (corresponding to \ang{;;0.9}), oriented perpendicular to the dispersion direction of each exposure and centered on each extraction position. } 
\label{fig:SPT-CLJ0411-4819}
\end{figure*}

\begin{figure*}
\centering
\includegraphics[angle=0,width=1.0\linewidth]{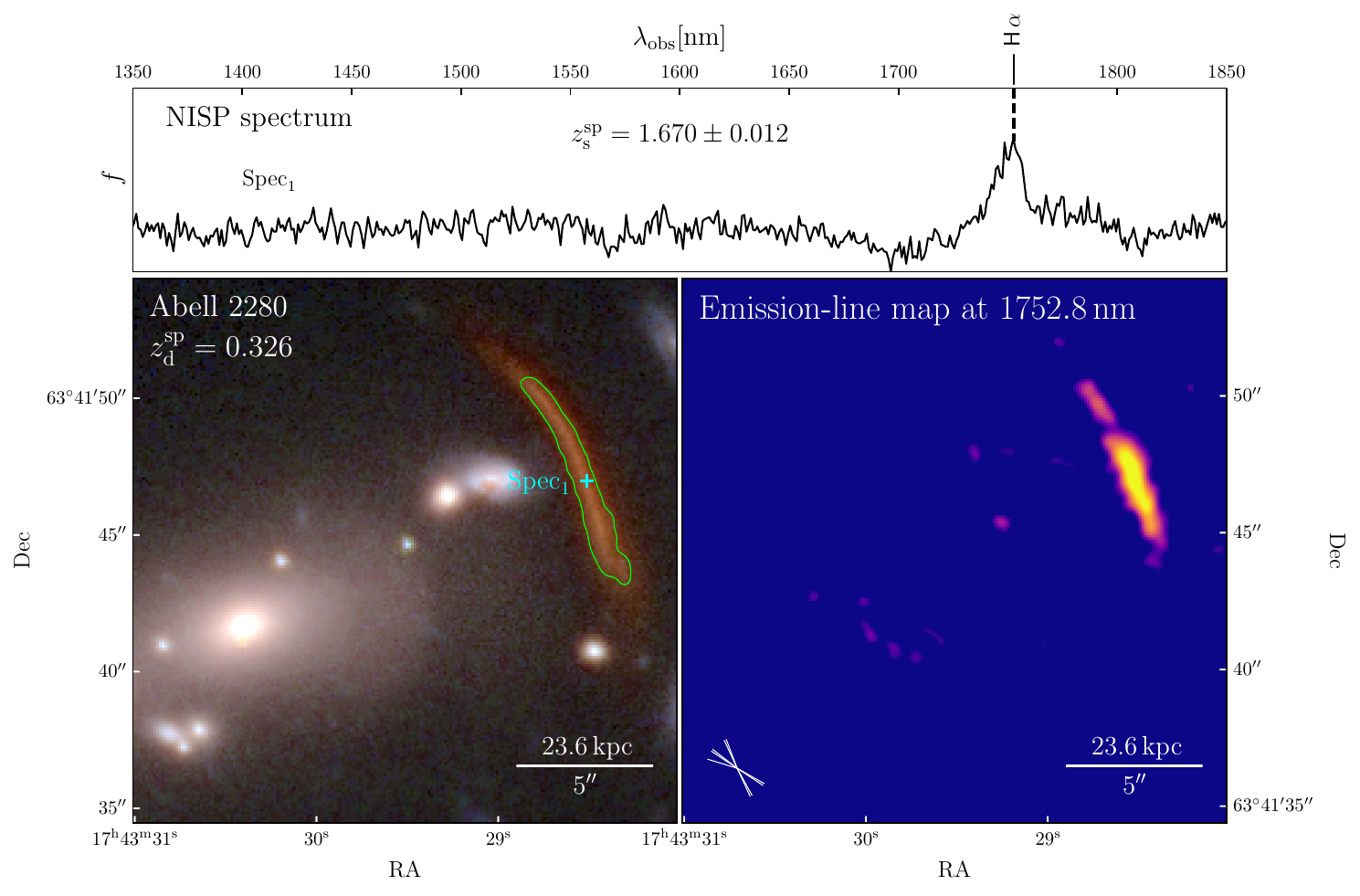}
\caption{Same as \Fig\ref{fig:SPT-CLJ0411-4819} for the galaxy cluster Abell~2280 at $z_\mathrm{d}^\mathrm{sp}=0.326\pm0.002$. A source with a measured \Euclid spectroscopic redshift of $z_\mathrm{s}^\mathrm{sp}=1.670\pm0.012$ is lensed into the red giant arc visible in the RGB cutout in the bottom-left panel. For this system, the top panel shows a single spectrum, extracted at the position marked by the cyan cross in the bottom-left panel and obtained by stacking nine \Euclid spectroscopic exposures.}
\label{fig:Abell2280}
\end{figure*}

\begin{figure*}
\centering
\includegraphics[angle=0,width=1.0\linewidth]{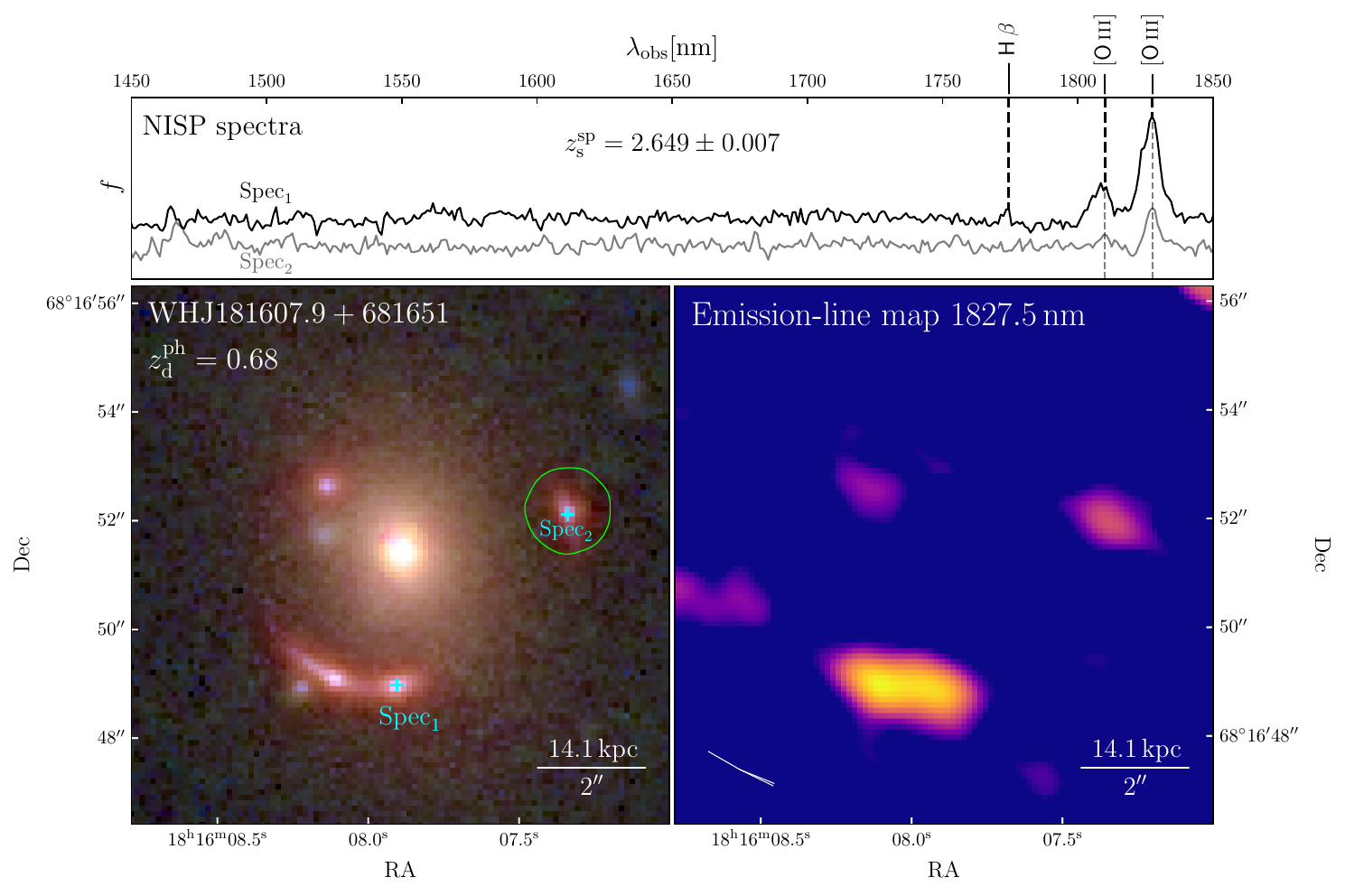}
\caption{Same as \Fig\ref{fig:SPT-CLJ0411-4819} for the WHJ181607.9$+$681651 system. The lens galaxy at $z_\mathrm{d}^\mathrm{ph}=0.68\pm0.04$, located at the centre of the RGB cutout in the bottom-left panel, produces four red multiple images of a background source with a measured \Euclid spectroscopic redshift of $z_\mathrm{s}^\mathrm{sp}=2.649\pm0.007$. Three emission lines, identified as the [\ion{O}{iii}] doublet and possibly \hb, are detected in the \Euclid ``$\mathrm{Spec_1}$'' spectrum of the lensed source (black 1D-spectrum in the top panel). Only the stronger component of the [\ion{O}{iii}] doublet is detected in ``$\mathrm{Spec_2}$'', displayed in light grey, at a lower signal-to-noise ratio. Both spectra were obtained by stacking three \Euclid spectroscopic exposures. The [\ion{O}{iii}] emission from the four multiple images is visible in the emission-line map.}
\label{fig:J181607}
\end{figure*}

\begin{figure*}
\centering
\includegraphics[angle=0,width=1.0\linewidth]{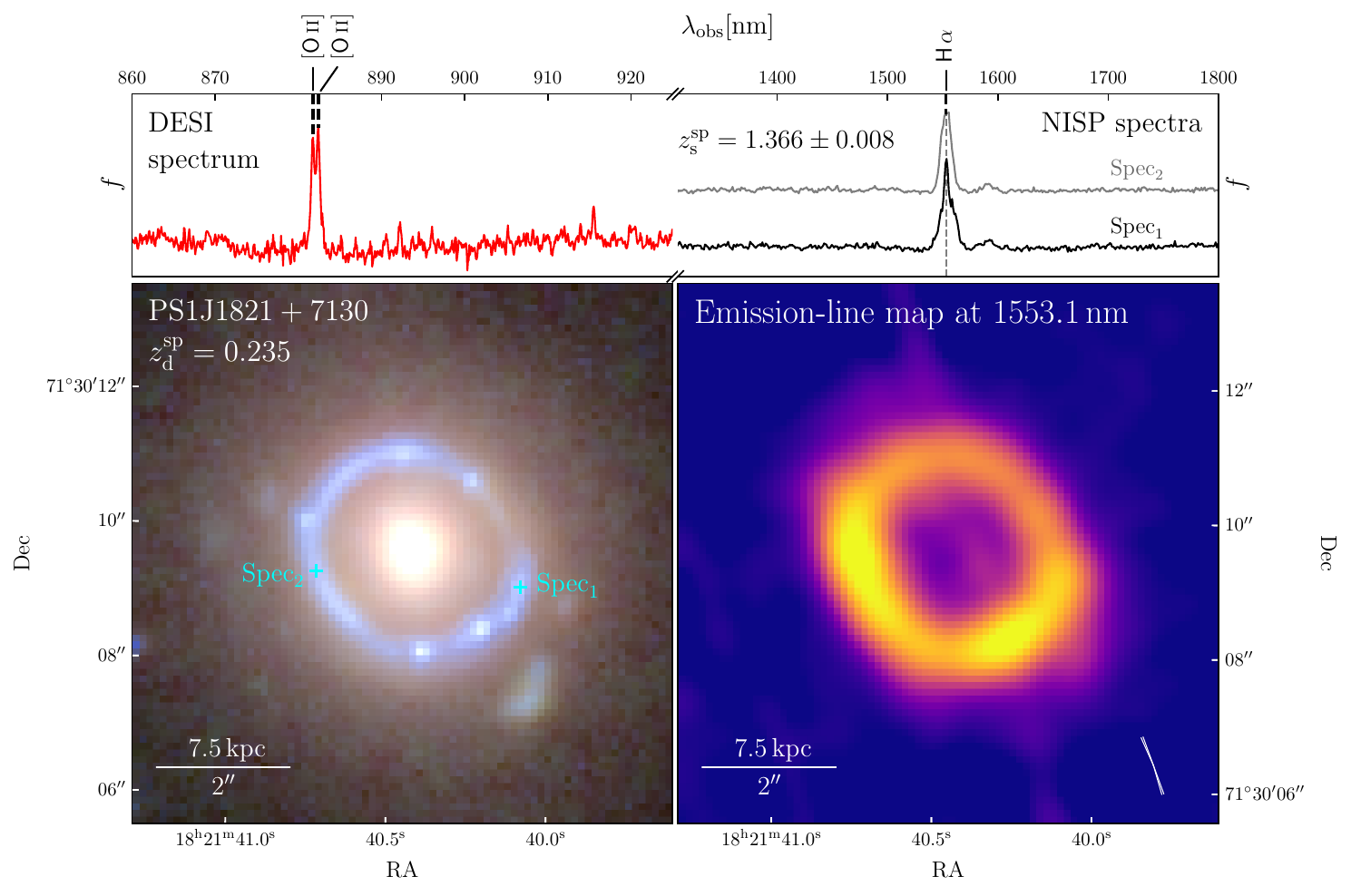}
\caption{\Euclid data of the galaxy-scale strong lensing system PS1J1821+7130 at $z_\mathrm{d}^\mathrm{sp}=0.235\pm0.001$. {\it Bottom-left:} RGB image of the system obtained by combining the \Euclid photometric bands (\HE, \YE, \JE, and \IE) using the \texttt{azulero} code \citep{Basset_Azulero}. {\it Bottom-right:} Emission-line map at 1553.1\,nm, showing the \ha emission associated with the Einstein ring. The map is obtained by combining four \Euclid dispersed spectroscopic images, whose corresponding dispersion directions are indicated by the white segments in the bottom-right corner of the panel. The map, oversampled at the native pixel scale of the \IE filter (\ang{;;0.1}), is smoothed with a Gaussian kernel with a standard deviation of two pixels (\ang{;;0.2}). {\it Top-right:} Stacked \Euclid NISP spectra (in arbitrary units), extracted at the positions marked by the two cyan crosses in the bottom-left panel. The \ha emission line from the Einstein ring is clearly detected in both spectra, yielding a \Euclid spectroscopic redshift for the lensed source of $z_\mathrm{s}^\mathrm{sp}=1.366\pm0.008$. {\it Top-left:} DESI spectrum highlighting the [\ion{O}{ii}] doublet emission lines at wavelengths consistent with the redshift measured from the \Euclid data.}
\label{fig:PS1J1821+7130}
\end{figure*}

In addition to the spectroscopic data, we make use of \Euclid photometric observations in the \IE, \YE, \JE, and \HE bands, at the maximum depth available for DR1, to identify strong-lensing features and define optimal positions for spectral extraction. The colour images shown in Figs.\,\ref{fig:SPT-CLJ0411-4819}, \ref{fig:Abell2280}, \ref{fig:J181607}, and \ref{fig:PS1J1821+7130} are produced by combining these photometric bands using the publicly available \texttt{azulero} code developed by \citet{Basset_Azulero}. Finally, in Appendix\,\ref{sc:PhotometricRedshift}, the \Euclid photometric data, complemented with ground-based multi-band observations, are used to estimate the photometric redshifts of the lensed sources, thereby providing an independent validation of the NISP results.  

\begin{table*}
    \centering
    \caption{Celestial coordinates and redshifts of the analysed strong lensing systems.}
    \begin{tabular}{lrrccrr}
    \hline
    \hline
    \multirow{2}{*}{ID} & 
    RA [deg] & 
    Dec [deg] & 
    \multirow{2}{*}{$z_\mathrm{d}$} & 
    \multirow{2}{*}{$z_\mathrm{s}^\mathrm{sp}$} & \multirow{2}{*}{$N_\mathrm{exp}$} &
    \multirow{2}{*}{\HE} \\ 
    & (J2000) & (J2000) & & \\
    \hline
    \\
    SPT-CLJ0411$-$4819 & 62.79574 & $-$48.32765 & $0.424\pm0.006\,^\mathrm{(a)}$ & $1.771\pm0.006$ &
    38 & 
    19.7 \\[5pt]
    Abell~2280 & 265.87663 & 63.69491 & $0.326\pm0.002\,^\mathrm{(b)}$ & $1.670\pm0.012$ & 
    9 & 
    18.2 \\[5pt]
    WHJ181607.9$+$681651 & 274.03285 & 68.28095 & $0.68\pm0.04\,^\mathrm{(c)}$ & $2.649\pm0.007$ & 
    3 & 
    21.1 \\[5pt]
    PS1J1821+7130 & 275.41843 & 71.50267 & $0.235\pm0.001\,^\mathrm{(d)}$ & $1.366\pm0.008$ & 
    4 & 
    17.1 \\[5pt] 
    \hline
    \end{tabular}
    \tablefoot{The system IDs and celestial coordinates of the gravitational lens centres are listed in the first three columns. The fourth column reports the deflector redshifts, $z_\mathrm{d}$, from the literature, while the fifth column gives the spectroscopic redshifts of the lensed background sources, $z_\mathrm{s}^\mathrm{sp}$, derived from the \Euclid data. The sixth column lists the number of spectroscopic exposure, $N_\mathrm{exp}$, each with an integration time of 550 seconds, stacked to generate the emission-line maps shown in Figs.\,\ref{fig:SPT-CLJ0411-4819}, \ref{fig:Abell2280}, \ref{fig:J181607}, and \ref{fig:PS1J1821+7130}. The last column gives the \HE magnitudes of the lensed images, measured within the green apertures shown in the bottom-left panels of Figs.\,\ref{fig:SPT-CLJ0411-4819}, \ref{fig:Abell2280}, and \ref{fig:J181607}, respectively. For PS1J1821+7130, the quoted value corresponds to the integrated magnitude of the Einstein ring.

    \textbf{References.} 
    (a) Spectroscopic redshift from \cite{Bayliss_2016},
    (b) Spectroscopic redshift from \cite{Gioia_1995},
    (c) Photometric redshift from \cite{Wen_2024}, and 
    (d) Spectroscopic redshift from DESI DR1 \citep{DESI_2026}.
    }
    
    \label{tab:table_redshifts}
\end{table*}

\section{\label{sc:SpectralExtraction}Spectral extraction and emission-line maps}

The customised spectral extractions presented in this paper were carried out on the calibrated NISP-S frames produced by the SIR processing function in the \Euclid Science Ground Segment pipeline (Euclid Collaboration: Scodeggio et al., in prep.). 
We used the {\tt GELSA}\footnote{\url{https://github.com/elsa-euclid/gelsa}, developed as part of the \href{https://elsa-euclid.github.io/}{Euclid Legacy Science Advanced analysis tools (ELSA)} project.} {\tt Python} package \citep{DR1-Gabarra} to process the NISP frames and perform a nearly automated spectral extraction of strong lensing features. This analysis utilise the SIR calibration products to map the spectral traces on the detector and apply the appropriate flux calibration \citep{DR1-TP024}. {\tt GELSA} provides routines to resample the NISP-S dispersed images onto a wavelength grid, producing both two-dimensional (2D) and one-dimensional (1D) spectra. Contamination from overlapping continuum was removed from the individual exposures by applying a 1D median filter of width 100 pixels ($\sim$137\,nm) to each row of the dispersed image along the dispersion direction. In the case of tilted exposures, the image is rotated to align the dispersion direction with pixel rows. The result is an estimate of the continuum, which is then subtracted from the original dispersed image to isolate the emission-line signal. In slitless spectroscopy, the spectrogram is, to first order, the result of the convolution of the source image with the 1D spectrum, and the dispersion direction depends on the orientation of the telescope and grism configuration \citep{Neveu2024, Outini2020}. This leads to challenges in the interpretation of the 1D spectral extraction of an extended source, such as an arc \citep[see also][]{DR1-Gabarra}. These limitations can be overcome by extracting emission-line maps from slitless spectroscopy. {\tt GELSA} produces emission-line maps by resampling the dispersed images in right ascension and declination at fixed wavelength. Since the maps approximately preserve the source image on the sky, exposures with different dispersion directions can be stacked to increase the ${\rm S/N}$. {\tt GELSA} resamples and combines the maps from different grism exposures using the drizzle algorithm \citep{Drizzle}. Emission-line maps from slitless spectroscopy differ from narrow-band imaging or integral-field spectroscopy. Emission-line maps are not truly monochromatic; the emission line is isolated by applying continuum subtraction on the 2D dispersed image, but emission from neighbouring lines such as \ha and [\ion{N}{ii}] that are displaced along the dispersion direction are superimposed on the map. 

\section{\label{sc:Results}Results}

In Figs.\,\ref{fig:SPT-CLJ0411-4819}, \ref{fig:Abell2280}, \ref{fig:J181607}, and \ref{fig:PS1J1821+7130}, we present the results of the analysis of the \Euclid slitless spectroscopic data for the strong lensing features considered in this work. The spectral extractions are performed at the positions marked by cyan crosses in the bottom-left panels of the respective figures. For SPT-CLJ0411$-$4819, WHJ181607.9$+$681651, and PS1J1821+7130, these positions correspond to two high-fidelity candidate counterimages of the same background galaxy or, in the case of PS1J1821+7130, to two opposite sides of the Einstein ring. For Abell~2280, instead, a single extraction position located on the brighter portion of the giant red arc is adopted. The spectra are extracted by using linear spatial apertures oriented perpendicular to the respective dispersion directions and centered on each selected extraction position. The aperture widths are set to three NISP pixels, corresponding to \ang{;;0.9}, for all systems.

We note that, unlike integral-field and slit spectroscopy, slitless spectroscopy is affected by the source morphology and extent along the dispersion direction, which affects the observed shape of spectral features in the extracted grism spectra. As a result, the spatial extent of the line-emitting region projected along this direction introduces additional morphological broadening beyond that due to the effective spectral resolution of the NISP-S grisms. Furthermore, the extraction positions are carefully selected to minimize spectral contamination from lensed counterimages. Specifically, we ensure that the dispersion traces do not intersect other counterimages whose emission lines could otherwise contaminate the extracted spectra at different positions along the spectral axis.

The redshifts of the background lensed sources are measured from the peak wavelength of the most prominent detected emission line of the ``$\mathrm{Spec_1}$’’ spectrum, if such a line can be uniquely identified. The quoted redshift uncertainties are derived from the standard deviation of the best-fitting Gaussian profiles to such emission line. We stress that this approach yields redshift uncertainties that are affected by the morphological broadening inherent to slitless spectroscopy, in addition to the instrumental broadening. In general, when the arc is visible in individual grism exposures, the redshift can be determined precisely during the construction of the emission-line map through the alignment of different grism orientations accounting for the source morphology. We note that forward-modelling approaches can further improve the redshift determination. Such an analysis is currently under development and is beyond the scope of this proof-of-concept paper.

The stacked 1D spectra of Abell~2280 and PS1J1821$+$7130, displayed in the top panels of Figs.\,\ref{fig:Abell2280} and \ref{fig:PS1J1821+7130}, respectively, each show a single prominent emission line. For PS1J1821$+$7130, the same spectral feature is clearly detected in the spectra independently extracted at the ``$\mathrm{Spec_1}$'' and ``$\mathrm{Spec_2}$'' positions. Assuming that these emission lines correspond to \ha emission, we measure spectroscopic redshifts of $1.670\pm0.012$ and $1.366\pm0.008$ for the lensed background sources in the Abell~2280 and PS1J1821+7130 systems, respectively.

Conversely, the stacked \Euclid spectra of the background galaxy lensed into the giant gravitational arc of the SPT-CLJ0411$-$4819 system, displayed in the top panel of \Fig\ref{fig:SPT-CLJ0411-4819}, reveal at least three emission lines at 1819.1\,nm, 1387.8\,nm, and 1347.4\,nm (the last two being barely visible in the spectrum extracted at the ``$\mathrm{Spec_2}$'' position). These spectral features are identified as \ha, [\ion{O}{iii}]\,$\lambda500.8$\,nm, and \hb emission lines, yielding a spectroscopic redshift of $z_\mathrm{s}^\mathrm{sp}=1.771\pm0.006$ for the lensed galaxy.

Finally, the \Euclid spectrum of the quadruply imaged lensed source in the WHJ181607.9$+$681651 system extracted at the ``$\mathrm{Spec_1}$'' position (\Fig\ref{fig:J181607}) contains two distinct emission lines at 1810.0\,nm and 1827.5\,nm. In addition, a third, fainter emission line is tentatively detected at 1774.4\,nm. The wavelengths and relative separations of these lines are consistent with the [\ion{O}{iii}] doublet and \hb emission from a lensed source at $z_\mathrm{s}^\mathrm{sp}=2.649\pm0.007$. 
We note that the strong [\ion{O}{iii}] emission relative to \hb suggests the presence of an AGN, consistent with its morphology (\Sec\ref{sc:WHJ181607.9+681651}). 
We also note that, at the ``$\mathrm{Spec_2}$'' position, corresponding to a less magnified counterimage of the background source than the one associated with ``$\mathrm{Spec_1}$'', only the stronger component of the [\ion{O}{iii}] doublet is detected.

In the bottom-right panels of Figs.\,\ref{fig:SPT-CLJ0411-4819}, \ref{fig:Abell2280}, \ref{fig:J181607}, and \ref{fig:PS1J1821+7130}, we show emission-line map images generated at the wavelength of the most prominent detected emission line for each system. In all cases, the morphology of the line-emitting regions closely reproduces that of the observed strong lensing features. The detection of emission at the same wavelength across multiple images, together with the close agreement in morphology and spatial extent with their observed counterparts, provides robust confirmation of the reliability of the detected emission lines and rules out the possibility that they arise from unrelated sources contaminating the dispersed spectra. At the same time, these findings provide further evidence that the multiple images originate from the same lensed background source.

To further corroborate the reliability of the \Euclid spectroscopic redshift measurements, which for two of the four systems are based on the identification of a single emission line, in Appendix\,\ref{sc:PhotometricRedshift} we compare the  \Euclid-NISP redshift values with photometric redshift estimates. We note that the PS1J1821$+$7130 system is not considered in this analysis since, as outlined below, DESI provides an independent spectroscopic confirmation of the \Euclid redshift. The photometric redshift estimates for SPT-CLJ0411$-$4819 and WHJ181607.9$+$681651 are consistent, within $1\sigma$, with the NISP-based spectroscopic redshift measurements. For Abell~2280, the spectroscopic redshift lies on the edge of the $1\sigma$ confidence interval of the photometric redshift posterior distribution. Overall, these results provide an independent confirmation of the reliability of the NISP-based spectroscopic redshifts. We refer the reader to Appendix\,\ref{sc:PhotometricRedshift} for further details.
 
As anticipated, for the PS1J1821$+$7130 system, a more robust validation of the \Euclid spectroscopic redshift value is provided by DESI DR1 data. Indeed, the compactness of this lensing system, combined with the diameter of the DESI optical fibers ($\sim\ang{;;1.5}$ on the sky) and the typical Kitt Peak point-spread function (PSF, $\sim\ang{;;1.1}$), causes part of the emission from the Einstein ring to fall within the DESI fiber centered on the lens galaxy. As a result, two clear emission lines, corresponding to the [\ion{O}{ii}] doublet, are visible in the DESI spectrum at wavelengths that closely match those expected from the redshift based on the identification of the \ha line from \Euclid spectroscopy (see the top panel of \Fig\ref{fig:PS1J1821+7130}).

As a simple lensing application of the aforementioned \Euclid redshifts (Table\,\ref{tab:table_redshifts}), we can derive a preliminary rough estimate of the total mass, $M(\theta_\sfont{E})$, enclosed within the Einstein angle, $\theta_\sfont{E}$, of each gravitational lens:
\begin{equation}
    M(\theta_\sfont{E}) = \frac{c^2}{4 G_\sfont{N}}\frac{\Dd \Ds}{\Dds}\theta_\sfont{E}^2\;.
    \label{eq:masslensing}
\end{equation}
Here, $G_\sfont{N}$ is the Newton’s gravitational constant, $c$ is the speed of light in vacuum, and \Dd, \Ds, and \Dds\ denote the angular-diameter distances to the lens, to the source, and between the lens and the source, respectively. These quantities can be directly derived from the measured spectroscopic redshifts, while an approximate estimate of $\theta_\sfont{E}$ for each strong lensing system can be obtained by measuring the projected distance between the lens centre (reported in Table\,\ref{tab:table_redshifts}) and the observed strong lensing features. 

Specifically, by measuring the angular separation between the center of the deflector and the lensed images, we estimate Einstein radii of approximately \ang{;;7.5}, \ang{;;13.5}, \ang{;;2.5} (as also reported by \citealt{Q1-SP048}), and \ang{;;1.5} \citep{DR1-SL03} for SPT-CLJ0411$-$4819, Abell~2280, WHJ181607.9$+$681651, and PS1J1821+7130, respectively. From these rough estimates of the Einstein radii, we derive $M(\theta_\sfont{E})$ values of $\qty{1e13}{\solarmass}$ and $\qty{3e13}{\solarmass}$ for the cluster-scale lenses SPT-CLJ0411$-$4819 and Abell~2280, respectively. These enclosed masses correspond to $2\text{--}6\%$ of the $M_\mathrm{500c}$ values of the clusters (Sects.\,\ref{sc:SPT-CLJ0411-4819} and \ref{sc:Abell2280}). This is expected, as the mass within the Einstein radius represents only a small fraction of the total cluster mass, which extends to much larger radii \citep[e.g.,][]{Fox_2022, Biviano2023ApJ...958..148B}.

For the bright lens galaxy in the WHJ181607.9$+$681651 system, we infer a mass of $\qty{2e12}{\solarmass}$. Finally, the lens galaxy in the PS1J1821+7130 system has an estimated mass of the order of $\qty{3e11}{\solarmass}$. Robust mass measurements will require detailed strong lensing models that account for the mass distribution of the deflectors and their environments. The development of such models is beyond the scope of this paper and is deferred to future work. 

\section{\label{sc:Conclusions}Conclusions}

Spectroscopic redshift measurements for multiple images and giant arcs in strong lensing systems are an indispensable ingredient for developing high-precision strong lensing models of galaxies, galaxy groups, and galaxy clusters. In this context, we present one of the first spectroscopic redshift determinations for lensed sources based on data obtained with the \Euclid NISP slitless spectrograph. Specifically, our analysis focused on two cluster-scale (SPT-CLJ0411$-$4819 and Abell 2280) and two galaxy-scale (WHJ181607.9$+$681651 and PS1J1821+7130) strong lensing systems. The \Euclid spectra of the gravitational arcs in Abell 2280 and PS1J1821+7130 show a single emission line, which we identified as the \ha transition. Under this assumption, we measured a redshift of $z_\mathrm{s}^\mathrm{sp}=1.670\pm0.012$ and $z_\mathrm{s}^\mathrm{sp}=1.366\pm0.008$, respectively. In the NISP spectra of the giant arc in the SPT-CLJ0411$-$4819 system, we identified three emission lines corresponding to \ha, [\ion{O}{iii}]\,$\lambda500.8$\,nm, and \hb transitions, which allowed us to securely determine the redshift of the background source as $z_\mathrm{s}^\mathrm{sp}=1.771\pm0.006$.
In the spectrum of the remaining system, WHJ181607.9$+$681651, we identified two emission lines that we associated with the [\ion{O}{iii}] doublet, together with a tentative detection of \hb, yielding a spectroscopic redshift of $z_\mathrm{s}^\mathrm{ph}=2.649\pm0.007$. 

The measured spectroscopic redshifts of SPT-CLJ0411$-$4819, Abell 2280, and WHJ181607.9$+$681651 are consistent with the photometric redshifts derived from \Euclid \IE, \YE, \JE, and \HE data, complemented by ground-based Hyper Suprime-Cam (HSC), MegaCam, and Dark Energy Camera (DECam) observations (Appendix\,\ref{sc:PhotometricRedshift}). As for PS1J1821+7130, a strong validation of the \Euclid spectroscopic redshift of the prominent Einstein ring, based on the identification of the \ha line at $z_\mathrm{s}^\mathrm{sp}\simeq 1.366$, was obtained from the DESI optical spectrum.

From the NISP slitless spectroscopic data, we generated continuum-subtracted emission-line maps at the wavelength of the most prominent emission line detected in each system. The features observed in these maps closely match the morphology of the lensed images in the VIS/NISP photometric data, further supporting the robustness of our results. In addition, such a morphological and positional match among emission line features demonstrates that the observed extended arcs and multiple images originates from the same background source, thus ruling out the hypothesis that the detected emission arises from contamination by unrelated sources. These results also highlight the potential of using collections of emission-line maps generated over a range of wavelengths as a powerful method for identifying counterimages in strong-lensing systems revealed by \Euclid data.

The present work represents an important milestone for future studies that will exploit \Euclid spectroscopic data to measure redshifts of strongly lensed giant arcs and multiple images. In addition, it provides both a methodology and a presentation of dedicated tools to analyse such data. In the near future, we plan to apply a similar methodology to the large sample of strong gravitational lenses discovered in the Euclid Q1 and DR1 datasets. A sizeable fraction of these systems are expected to exhibit detectable emission lines within the NISP wavelength range, thereby providing the critical spectroscopic information required for robust strong-lensing analyses. 
Lastly, this study paves the way for the development of the first strong lensing models based entirely on \Euclid spectrophotometric observations.

\begin{acknowledgements}
MB, MM, and PB were supported by ASI Grant n. 2024-10-HH.0 "Attività scientifiche per la missione Euclid – fase E". MM and PB also acknowledge financial support from the INAF Grants “The Big-Data era of cluster lensing" and "Probing Dark Matter and Galaxy Formation in Galaxy Clusters through Strong Gravitational Lensing", as well as from PRIN-MIUR 2020SKSTHZ and 20229YBSAN. MB acknowledges the support from INAF Minigrant 2023 ‘ADIEU’. GG and NL acknowledge travel support by Euclid:UK through STFC grant APP91126. LT acknowledges fundings from the Deutsche Forschungsgemeinschaft (DFG, German Research Foundation) under Germany’s Excellence Strategy – EXC-2094 – 390783311. CJS acknowledges funding from the NASA Euclid General Investigator Program Grant No. 80NSSC25K7735. AA acknowledges financial support through the Beatriz Galindo programme and the project PID2025-168678NB-I00 (MCIU/AEI/MINECO/FEDER, UE), Ministerio de Ciencia, Investigación y Universidades. The Instituto de Física de Cantabria (IFCA) has been accredited as Unidad de Excelencia María de Maeztu, a distinction awarded by the Spanish State Research Agency for the period 01/06/2026 - 31/05/2030. LB acknowledges the support by the Italian Ministry of University and Research (grant FIS2023-01611, CUP C53C25000300001). JWN is supported by an STFC/UKRI Ernest Rutherford Fellowship, Project Reference: ST/X003086/1. NL is supported through a graduate studentship from the UKRI STFC and the University of Portsmouth, and acknowledges travel support provided by STFC for UK participation in the Rubin LSST through grant ST/X001334/1.

The authors acknowledge support from the ELSA project. "ELSA: Euclid Legacy Science Advanced analysis tools" (Grant Agreement no. 101135203) is funded by the European Union. UK participation is funded through the UK Horizon guarantee scheme under Innovate UK grant 10093177. Views and opinions expressed are however those of the author(s) only and do not necessarily reflect those of the European Union or Innovate UK. Neither the European Union nor the granting authority can be held responsible for them. 
  
  \AckDRone
  \AckDatalabs
  \AckEC 

This research used data obtained with the Dark Energy Spectroscopic Instrument (DESI). DESI construction and operations is managed by the Lawrence Berkeley National Laboratory. This material is based upon work supported by the U.S. Department of Energy, Office of Science, Office of High-Energy Physics, under Contract No. DE–AC02–05CH11231, and by the National Energy Research Scientific Computing Center, a DOE Office of Science User Facility under the same contract. Additional support for DESI was provided by the U.S. National Science Foundation (NSF), Division of Astronomical Sciences under Contract No. AST-0950945 to the NSF’s National Optical-Infrared Astronomy Research Laboratory; the Science and Technology Facilities Council of the United Kingdom; the Gordon and Betty Moore Foundation; the Heising-Simons Foundation; the French Alternative Energies and Atomic Energy Commission (CEA); the National Council of Humanities, Science and Technology of Mexico (CONAHCYT); the Ministry of Science and Innovation of Spain (MICINN), and by the DESI Member Institutions: www.desi.lbl.gov/collaborating-institutions. The DESI collaboration is honored to be permitted to conduct scientific research on I’oligam Du’ag (Kitt Peak), a mountain with particular significance to the Tohono O’odham Nation. Any opinions, findings, and conclusions or recommendations expressed in this material are those of the author(s) and do not necessarily reflect the views of the U.S. National Science Foundation, the U.S. Department of Energy, or any of the listed funding agencies.
\end{acknowledgements}

\bibliography{Euclid, bibliography}

\begin{appendix}

\begin{strip}
\centering
\renewcommand{\thefigure}{A.1}
\includegraphics[angle=0,width=1.0\linewidth]{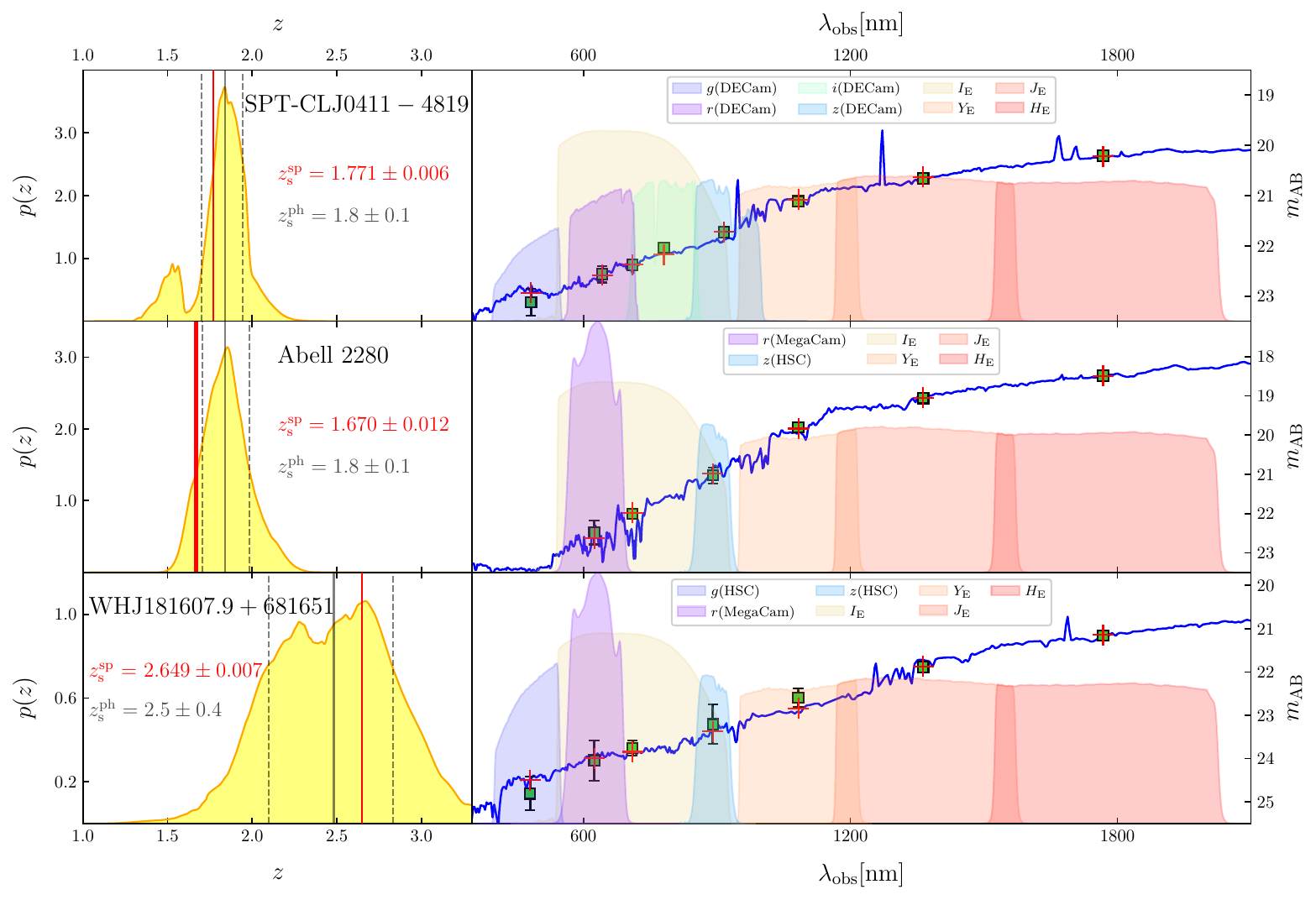}
\captionof{figure}{{\it Left panels:} Posterior probability distributions of the photometric-redshift estimates, corrected for systematic errors. The vertical grey lines mark the 50th (solid) and 16th and 84th (dashed) percentiles of the distributions, while the vertical red lines indicate the spectroscopic redshifts of the lensed background sources, as inferred from the emission lines detected in the \Euclid spectra (Table\,\ref{tab:table_redshifts}). The widths of the red lines represent the uncertainties associated with the spectroscopic redshift measurements. The measured spectroscopic and photometric redshift values are also reported in red and grey text, respectively. {\it Right panels:} Spectral energy distributions (SEDs) of the lensed sources. Green squares indicate the AB magnitudes measured from \Euclid and ground-based imaging data, while the overplotted coloured regions show the transmission curves of the corresponding filters. Magnitudes and their associated uncertainties are measured within the green apertures shown in the bottom-left panels of Figs.\,\ref{fig:SPT-CLJ0411-4819}, \ref{fig:Abell2280}, and \ref{fig:J181607}, after PSF matching. The blue curves show the best-fitting SED templates obtained with \texttt{eazy-py}, while the red crosses indicate the corresponding model-predicted photometry.}
\label{fig:appendix}
\end{strip}

\section{\label{sc:PhotometricRedshift}Photometric redshift analysis}

The \Euclid slitless data often enable the detection of only a single emission line over the NISP wavelength range, as in the case of two of the four analysed targets. As a result, the redshift determination remains uncertain due to the ambiguity in associating  the detected line with its corresponding atomic transition. This difficulty can be overcome by extending the spectral coverage, typically in the optical regime, with complementary observations (see the case of PS1J1821+7130), or with the aid of independent photometric redshift estimates. 

In this appendix, we derive photometric redshift estimates for the lensed galaxies in the SPT-CLJ0411$-$4819, Abell~2280, and WHJ181607.9$+$681651 systems to independently corroborate the NISP spectroscopic results. To this aim, 
the \Euclid photometric data in the \IE, \YE, \JE, and \HE bands are complemented by optical imaging from ground-based facilities available through the \Euclid archive. These ground-based observations are coordinated by the \Euclid organisation unit for external data (OU-EXT). In particular, coverage of the northern hemisphere is provided by the Ultraviolet Near Infrared Optical Northern Survey \citep[UNIONS,][]{UNIONS_2025}, whereas coverage in the southern hemisphere is currently provided by DES \citep{DES_2016}. In addition, the EDFs are further complemented by dedicated observing campaigns and archival observations \citep{DR1-TP006}.
Specifically, for SPT-CLJ0411$-$4819, we used observations in the $g$, $r$, $i$, and $z$ filters obtained with DECam on the Blanco Telescope. For Abell~2280, we used $r$-band observations obtained with the MegaCam wide-field optical imager on the CFHT, together with $z$-band imaging from the HSC on the Subaru Telescope. Finally, for WHJ181607.9$+$681651, we used HSC observations in the $g$ and $z$ filters and MegaCam $r$-band imaging (\Fig\ref{fig:appendix}). 

Although the \Euclid pipeline has proven to be generally effective in measuring photometric redshifts for field galaxies, the automated photometric analysis of strong lensing features produced by galaxy-, group- and cluster-scale lenses remains extremely challenging. Due to the small angular separation between the lens galaxy and the lensed images, as well as the crowding in cluster and group fields, most of the strong lensing features are not automatically deblended by the pipeline. In addition, the distorted morphologies of the lensed galaxies require a supervised definition of the photometric apertures. For these reasons, we performed the photometric measurements using a customised approach tailored to the systems studied here. To perform PSF-matched photometric measurements of the strong lensing features, the angular resolution in each imaging dataset was degraded to match that of the lowest-resolution filter, corresponding to the DECam $g$ band for SPT-CLJ0411$-$4819 and the MegaCam $r$ band for Abell~2280 and WHJ181607.9$+$681651. This step ensures that photometric measurements are performed over the same physical area for each source. The right panels of \Fig\ref{fig:appendix} show the spectral energy distributions (SEDs) of the individual lensed features. The green squares represent the magnitudes and their associated uncertainties measured for each telescope/filter combination within the green apertures shown in the bottom-left panels of Figs.\,\ref{fig:SPT-CLJ0411-4819}, \ref{fig:Abell2280}, and \ref{fig:J181607}.

Photometric redshift estimates of lensed sources are obtained with the publicly available software \texttt{eazy-py} \citep{Brammer_2008} by fitting each SED with the set of 11 galaxy templates from \cite{Benitez_2000}, which span from starburst to early-type galaxies. 
To assess the robustness of the photometric redshifts derived above and quantify their statistical and systematic uncertainties, we used a control sample of 160 galaxies with spectroscopic redshifts in the range $0.3\lesssim z\lesssim 2.5$ from the DESI database. For this sample, we retrieved \Euclid and associated ground-based imaging data from the \Euclid archive, applied the same photometric procedure used for the lensed sources to construct their SEDs, and derived photometric redshift estimates using the same \texttt{eazy-py} configuration. By comparing photometric and spectroscopic redshifts of the 160 galaxies, we estimated a statistical uncertainty in the photometric redshifts of $\sigma_z = \ave{z_{\rm phot}-z_{\rm spec}}(1+z_{\rm spec})^{-1}=0.1$. However, photometric redshifts appear to be systematically underestimated, with an offset of $\Delta z_{\rm syst}=z_{\rm spec}-z_{\rm phot}\simeq 0.3$, for galaxies in the redshift range $1<z<2.5$, which is most relevant to our analysis. The latter is likely due to crowding in ground-based data, which makes it difficult to properly separate fluxes of close sources, and PSF matching when combining ground- and space-based imaging with very different PSFs. The posterior probability distributions of the photometric redshifts for each system are shown in the left-hand panels of Fig.\,\ref{fig:appendix}.
From this analysis, we obtained photometric redshift estimates of $1.8\pm0.1$, $1.8\pm0.1$, and $2.5\pm0.4$ for the SPT-CLJ0411$-$4819, Abell~2280, and WHJ181607.9$+$681651 systems, respectively. The redshift estimates were derived from the medians of the posterior distributions, while the corresponding uncertainties were determined from the 16th and 84th percentiles. These values are consistent within the 1$\sigma$ confidence level with the measured \Euclid spectroscopic redshift (reported in red in \Fig\ref{fig:appendix}), except for Abell~2280, for which we find a difference of $\Delta z=z_\mathrm{s}^\mathrm{ph}-z_\mathrm{s}^\mathrm{sp}\sim0.13$, on the edge of the 1$\sigma$ interval. We note that this difference is significantly smaller than the typical combined statistical and systematic uncertainties ($\sim0.4$) estimated from the control sample.

\end{appendix}

\label{LastPage}
\end{document}